%% file: PRA_2024.tex
\pdfoutput=1
\documentclass[%
reprint,
nofootinbib,
amsmath,amssymb,
aps,
pra,
]{revtex4-2}

\usepackage{graphicx}
\usepackage{dcolumn}
\usepackage{bm}

\usepackage{multirow}%
\usepackage{amsfonts}%
\usepackage{amsthm}%
\usepackage{mathrsfs}%
\usepackage{xcolor}%
\usepackage{algorithm}%
\usepackage{mathtools}
\mathtoolsset{showonlyrefs}
\usepackage{makecell}

\usepackage{nicematrix}
\usepackage{tikz}
\usepackage{pgfplots}
\pgfplotsset{compat=newest}
\usepackage{ifthen}
\usetikzlibrary{patterns, patterns.meta}
\usetikzlibrary{arrows.meta, shapes.geometric, decorations.pathmorphing, backgrounds, positioning, fit, petri, calc}
\usepackage{pgffor}
\usetikzlibrary{plotmarks}
\usepgfplotslibrary{patchplots}

\pgfdeclarepattern{
  name=hatch,
  parameters={\hatchsize,\hatchangle,\hatchlinewidth},
  bottom left={\pgfpoint{-.1pt}{-.1pt}},
  top right={\pgfpoint{\hatchsize+.1pt}{\hatchsize+.1pt}},
  tile size={\pgfpoint{\hatchsize}{\hatchsize}},
  tile transformation={\pgftransformrotate{\hatchangle}},
  code={
    \pgfsetlinewidth{\hatchlinewidth}
    \pgfpathmoveto{\pgfpoint{-.1pt}{-.1pt}}
    \pgfpathlineto{\pgfpoint{\hatchsize+.1pt}{\hatchsize+.1pt}}
    \pgfpathmoveto{\pgfpoint{-.1pt}{\hatchsize+.1pt}}
    \pgfpathlineto{\pgfpoint{\hatchsize+.1pt}{-.1pt}}
    \pgfusepath{stroke}
  }
}

\tikzset{
  hatch size/.store in=\hatchsize,
  hatch angle/.store in=\hatchangle,
  hatch line width/.store in=\hatchlinewidth,
  hatch size=5pt,
  hatch angle=0pt,
  hatch line width=.5pt,
}

\newtheorem{theorem}{Theorem}
\newtheorem{proposition}[theorem]{Proposition}%
\newtheorem{example}{Example}%

\begin{document}
\title{A High-Performance List Decoding Algorithm for Surface Codes with Erroneous Syndrome}


\author{Jifan Liang}
\affiliation{School of Computer Science and Engineering, Sun Yat-sen University, Guangzhou 510006, China}

\author{Qianfan Wang}
\affiliation{School of Computer Science and Engineering, Sun Yat-sen University, Guangzhou 510006, China}

\author{Lvzhou Li}
\affiliation{School of Computer Science and Engineering, Sun Yat-sen University, Guangzhou 510006, China}

\author{Xiao Ma}
\affiliation{School of Computer Science and Engineering, Sun Yat-sen University, Guangzhou 510006, China}

\date{\today}

\begin{abstract}
  Quantum error-correcting codes~(QECCs) are necessary for fault-tolerant quantum computation. 
  Surface codes are a class of topological QECCs that have attracted significant attention due to their exceptional error-correcting capabilities and easy implementation.
  In the decoding process of surface codes, the syndromes are crucial for error correction, however, they are not always correctly measured. 
  Most of the existing decoding algorithms for surface codes need extra measurements to correct syndromes with errors, which implies a potential increase in inference complexity and decoding latency.
  In this paper, we propose a high-performance list decoding algorithm for surface codes with erroneous syndromes, where syndrome soft information is incorporated in the decoding, allowing qubits and syndrome to be recovered without needing extra measurements.
  Precisely, we first use belief propagation~(BP) decoding for pre-processing with syndrome soft information, followed by ordered statistics decoding~(OSD) for post-processing to list and recover both qubits and syndromes.
  Numerical results demonstrate that our proposed algorithm efficiently recovers erroneous syndromes and significantly improves the decoding performance of surface codes with erroneous syndromes compared to minimum-weight perfect matching~(MWPM), BP and original BP-OSD algorithms.
\end{abstract}


\maketitle

\section{Introduction}
\label{sec: intro}
Quantum states are highly vulnerable to decoherence and noise, which can compromise the accuracy of quantum computations. 
Consequently, quantum error-correcting codes~(QECCs) are essential for preserving and manipulating quantum information in such environments~\cite{gottesman1997stabilizer}. 
QECCs ensure the integrity of quantum states, enabling fault-tolerant quantum computation and paving the way for practical and scalable quantum computers with high fidelity~\cite{nielsen2010quantum}.

\subsection{Surface codes and their decoding algorithms}
Surface codes are a class of topological QECCs that have garnered significant attention in the field of quantum error correction due to their exceptional error-correcting capabilities and high fault-tolerant thresholds~\cite{dennis2002topological}.
These codes are obtained by mapping two-dimensional toric codes~\cite{kitaev2003fault} onto a plane. 
In this configuration, logical qubits correspond to paths that extend from one boundary of the lattice to the opposite boundary.
Measurement of specific stabilizer operators associated with the lattice can detect and correct errors.
Numerous decoding algorithms have been proposed to address quantum errors in surface codes, with one of the most popular categories being the minimum-weight perfect matching~(MWPM) decoder~\cite{kolmogorov2009blossom, fowler2013minimum}.
This type of decoder maps error patterns to graphical representations and searches for the minimum-weight matching in the graph to correct errors.
Another category of decoding algorithms is based on the belief propagation~(BP) algorithm, widely used in classical error correction to decode low-density parity-check codes.
Several BP-based decoding algorithms have been proposed for surface codes, including the 
the generalized BP decoder~\cite{old2023generalized}, the enhanced BP decoder~\cite{chytas2024enhanced}, the branch BP decoder~\cite{huang2023branch}.
Recently, a post-processing method called ordered statistics decoding~(OSD), specifically BP-OSD, has demonstrated a significant improvement in decoding performance over the conventional BP decoder~\cite{panteleev2021degenerate, liang2024BPLCOSD}.

\subsection{Motivation}
Measuring the syndrome is a critical step in decoding algorithms for QECCs. 
However, measurement errors are inevitable in practice, and the syndromes may be corrupted by measurement errors~\cite{meinerz2022scalable, nemec2023quantum}.
The aforementioned algorithms are not well-equipped to handle erroneous syndrome information, which may fail to correct errors effectively, leading to a significant decline in performance.
By finding matches on a 3-dimensional lattice~\cite{bonilla2021xzzx} constructed from multiple measurements, the MWPM decoder has recently been enhanced for the scenerio with erroneous syndrome.
Nonetheless, this enhanced algorithm may increase the latency and complexity due to the required extra measurements.
Thus, it is crucial to develop high-performance decoding algorithms that can manage incorrect syndromes without increasing the number of measurements, while preserving computational efficiency and delivering good decoding performance.

\subsection{Contributions}
In this paper, we propose a high-performance list decoding algorithm for surface codes with erroneous syndromes. 
The main contributions of our work are as follows.

In the pre-processing stage, we introduce the enhanced BP decoder, which incorporate soft information~(\textit{a priori} information) about the syndrome during BP decoding, allowing iterative updates to obtain soft information for both qubits and syndromes. 
Additionally, we use normalized min-sum~(NMS) to further improve performance.

In the post-processing stage, we use the output from the pre-processing to introduce a virtual codeword with an extended parity-check matrix, enabling us to list and recover both syndrome and qubit errors.

By combining these two stages, we propose an extended BP-OSD algorithm for practical scenarios with erroneous syndromes, capable of correcting error patterns in both qubits and syndromes. Unlike other algorithms that typically require extra measurements to address incorrect syndromes~\cite{dennis2002topological}, our proposed algorithm can decode and recover from incorrect syndromes without needing additional measurements.

Numerical results demonstrate that (1)~the proposed extended BP-OSD algorithm can effectively recover erroneous syndromes, and (2)~qubit errors can be corrected efficiently, given the enhanced syndrome accuracy, significantly outperforming classical MWPM, BP and the original BP-OSD algorithms.

\subsection{Organization}
The remainder of this paper is structured as follows.
Sec.~\ref{sec: prelim} provides an overview of surface codes, the MWPM decoder, the extended BP-OSD algorithm and the channel model.
In Sec.~\ref{sec: method}, we introduce our proposed list decoding algorithm for handling surface codes with erroneous syndromes.
Sec.~\ref{sec: simulation} presents numerical results demonstrating the effectiveness of the proposed algorithm.
Finally, Sec.~\ref{sec: conclusion} concludes the paper.

\section{Preliminaries}
\label{sec: prelim}

\subsection{Stabilizer codes and surface codes}
\label{subsec: stabilizer}
Usually, stabilizer code with parameters $[[n, k]]$ is defined by an Abelian subgroup $\mathcal{S}$ of the general Pauli group on $n$ qubits $\mathcal{G}_n$, which is a group of $n$-fold tensor product of Pauli matrices~($I$, $X$, $Y$ and $Z$ operators) together with multiplicative factors $\pm 1, \pm i$.
The stabilizer group can be generated by $m = n-k$ generators, whose common eigenspace with eigenvalue $+1$ is the code space.
It can be shown that this common eigenspace is a $2^k$-dimensional subspace of the $2^n$-dimensional Hilbert space~(the $n$-qubit space).
If the code has a minimum distance of $d$, it is often denoted as $[[n, k, d]]$. 
This notation indicates that arbitrary errors affecting up to $\lfloor (d-1)/2 \rfloor$ qubits can be corrected.
To correct qubit errors, the stabilizer generators are measured, and the measurement results, called the syndrome, are used to determine the recovery operation by decoding algorithms.
By convention, the syndrome is represented as a binary string with components denoted by $0$ and $1$, which correspond to the eigenvalues $+1$ and $-1$ of the stabilizer operators, respectively.

General stabilizer codes are more challenging to implement and decode compared to topological codes like surface codes.
A $[[2d^2 - 2d + 1, 1, d]]$ surface code is defined on a two-dimensional square lattice with side length $(d-1)$.
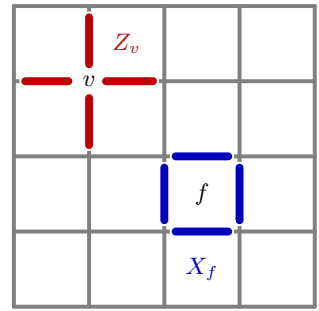
\begin{figure}[tb]
  \input{Figures/surface-stabilizer.tex}
  \caption{\label{fig: surface}The $[[n = 41, k = 1, d = 5]]$ surface code. In the lattice representation, $X$-type stabilizers are shown on the faces, while $Z$-type stabilizers are shown on the vertices.}
\end{figure}
The logical qubit is encoded into the lattice, whose edges are associated with physical qubits.
As shown in Fig.~\ref{fig: surface}, the surface code is defined by two types of generators: the $X$-type stabilizer associated with a face $f$ and the $Z$-type stabilizer associated with a vertex $v$.
For a face $f$ and a vertex $v$, the $X$- and $Z$-type stabilizers are formally defined as:
\begin{equation}
  X_f = \prod_{e\in \partial f} X_e \mbox{ and } Z_v = \prod_{e: v \in \partial e} Z_e, \, \mbox{respectively,}
\end{equation}
where $\partial f$ denotes the set of edges around face $f$, $\partial e$ denotes the set of endpoints of edge $e$, and $X_e$~($Z_e$) denotes the Pauli $X$~($Z$) operator on the physical qubit associated with edge $e$.
One can prove that all the $X$- and $Z$-type stabilizers commute with each other, i.e.,
for all faces $f, f'$ and vertices $v, v'$, 
\begin{equation}
    [X_f, X_{f'}] = 0, \, [X_f, Z_v] = 0, \, [Z_v, Z_{v'}] = 0, 
\end{equation}
where $[A, B] = AB - BA$ denotes the commutator of operators $A$ and $B$.
This commutativity is essential for stabilizer code, as it ensures that stabilizer operators can be measured simultaneously without interfering with each other~\cite{lidar2013quantum}.

\subsection{MWPM decoder}
\label{subsec: mwpm}
Since surface codes have planar~(graph) structures, graph algorithms can be applied to decode them.
One of the most popular decoding algorithms for surface codes is the MWPM decoder~\cite{kolmogorov2009blossom, fowler2013minimum}.
The MWPM decoder considers each face~(vertex) with a non-zero syndrome as a node in the graph, and the edges between nodes are usually weighted by the Manhattan distance between the faces~(vertices).
It pairs each node with another unmatched node or an artificial node~(representing the boundary of the lattice) and processes faces and vertices~(corresponding to $X$- and $Z$-stabilizers, respectively) separately.
The MWPM decoder aims to find a perfect matching with the minimum total weight, where `perfect' means that no nodes are left unmatched.
When the syndrome information is accurately measured, non-zero syndromes~(nodes) are only caused by pure Pauli errors~\cite{lidar2013quantum}, allowing the MWPM decoder to correct these errors with near-perfect accuracy.
Conversely, if the syndrome information is erroneous, the MWPM decoder needs to be modified, such as extending to a 3-dimensional lattice where the third dimension represents time, and requires multiple rounds of measurements~\cite{bonilla2021xzzx}.
This extension allows the MWPM decoder to correct syndromes with errors but may increase the latency and complexity of the decoder due to the extra measurements required.

\subsection{The BP-OSD algorithm}
\label{subsec: bp-osd}
Despite the elegance of the MWPM decoder, it still falls short of the performance of the optimal algorithm for surface codes, even with accurate syndrome information.
The BP-OSD algorithm, which combines the BP algorithm with the OSD algorithm, has been shown to be better at decoding surface codes~\cite{panteleev2021degenerate}.
The original BP algorithm~\cite{kschischang2001factor} is a message-passing algorithm and the original OSD algorithm~\cite{fossorier1995soft} is a list decoding algorithm, both of which are widely used to decode classical codes.
In the context of quantum error correction, these two classical decoding algorithms have been enhanced to address the short cycle and degenerate error problems~\cite{kuo2022exploiting}.

\subsection{System model}
\label{subsec: model}
The BP-OSD algorithm has been demonstrated to outperform both the MWPM and BP decoders in decoding surface codes with accurate syndromes~\cite{liang2024BPLCOSD}.
In this paper, we extend the BP-OSD algorithm to handle erroneous syndrome information, which is a common scenario in practice.
More specifically, we consider a depolarizing channel for qubits and a bit-flip channel for the measured syndrome.
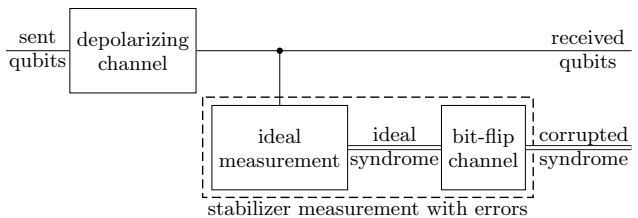
\begin{figure}[tb]
  \resizebox{.475\textwidth}{!}{\input{Figures/system-model.tex}}
  \caption{\label{fig: system}System model. The quantum information is sent through a depolarizing channel, and the ideal measured syndrome is corrupted by a bit-flip channel~(the single/double lines denote the quantum/classical information).}
\end{figure}
As illustrated in Fig.~\ref{fig: system}, the sent qubits subjected independent and identically distributed~(i.i.d.) depolarizing errors. Meanwhile, the measured syndrome is affected by i.i.d. bit-flip errors, meaning the measured syndrome is derived from the ideal (error-free) syndrome through a classical binary symmetric channel.
Moreover, the qubit and measurement errors are assumed to be not correlated with one another.

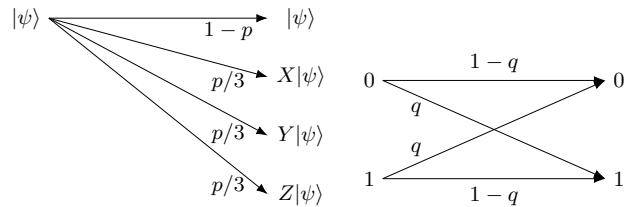
\begin{figure}[tb]
  \hfill
  \resizebox{.25\textwidth}{!}{\input{Figures/depolarizing-channel.tex}}
  \hfill
  \resizebox{.21\textwidth}{!}{\input{Figures/bit-flip-channel.tex}}
  \hfill
  \caption{\label{fig: error}The error model of the quantum channel with depolarizing rate $p$ and syndrome channel with bit-flip rate $q$.}
\end{figure}
the depolarizing rate of the quantum channel as $p$ and the bit-flip rate of the syndrome channel as $q$, the error model can be described as shown in Fig.~\ref{fig: error}, where $p$ and $q$ are assumed to be known.
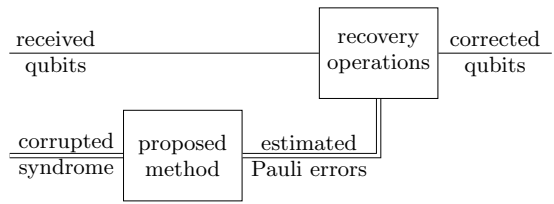
\begin{figure}[tb]
  \resizebox{.41\textwidth}{!}{\input{Figures/decoder-goal.tex}}
  \caption{\label{fig: procedure}The procedure of quantum error correction with erroneous syndrome~(the single/double lines denote the quantum/classical information).}
\end{figure}
Within this model, the main objective of the proposed method is to correct quantum channel errors and implicitly correct syndrome errors. The procedure is illustrated in Fig.~\ref{fig: procedure}.

\section{Proposed method}
\label{sec: method}
\begin{figure}[tb]
  \resizebox{.48\textwidth}{!}{\input{Figures/framework.tex}}
  \caption{\label{fig: framework}The framework of the proposed BP-OSD algorithm.}
\end{figure}
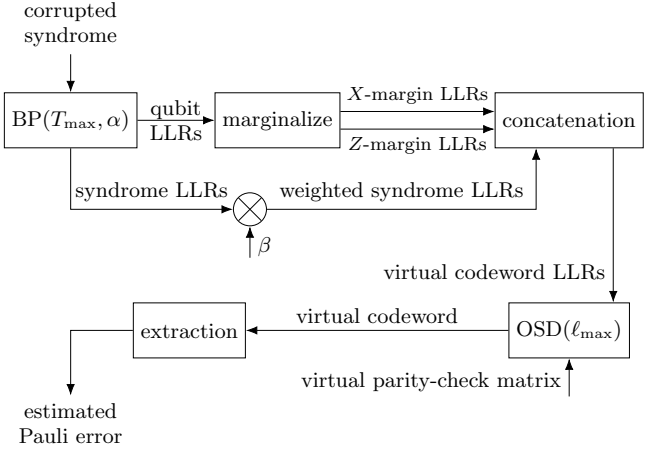

Aiming to correct both the qubit and the syndrome for the surface codes, we propose a high-performance list decoding algorithm that incorporates syndrome soft information, which we call the extended BP-OSD algorithm, or simply the BP-OSD algorithm.
Using the soft information of both qubits and syndromes, the OSD algorithm can naturally list and correct error patterns in both.
The proposed method consists of two main steps compared to the conventional BP-OSD algorithm~\cite{panteleev2021degenerate}.
\begin{enumerate}
  \item At the pre-processing stage, the \textit{a priori} soft information is introduced during the BP decoder, resulting in \textit{a posteriori} soft information for both the qubit and the syndrome.
  \item At the post-processing stage, the virtual codeword is introduced, and the OSD algorithm is applied to list and correct error patterns in both the qubit and the syndrome.
\end{enumerate}
The proposed BP-OSD algorithm is outlined in Fig.~\ref{fig: framework}, with detailed descriptions of each step being provided in the following subsections.

\subsection{Enhancing BP with \textit{a priori} syndrome information}
Since the OSD algorithm takes log-likelihood ratios~(LLRs) as input, we first use the BP algorithm to improve the accuracy of the LLRs for both qubits and syndrome. 
As usual, our BP decoder processes the quantum channel LLRs and outputs the LLRs for each qubit, just like the conventional quantum BP decoder. 
Instead of relying on hard syndrome information, we improve the BP decoder by incorporating syndrome soft information, referring to this as the enhanced BP decoder.
To clarify, we use the following example~(Example~\ref{ex: codes}) to illustrate our approach to initializing the soft information.
\begin{example}
  \label{ex: codes}
  The $[[5, 1, 3]]$ perfect stabilizer code~\cite{laflamme1996perfect} is the smallest quantum code capable of protecting a single qubit from any arbitrary single-qubit error.
  This code is defined by the stabilizer group
  \begin{equation}
    \mathcal{S} = \langle XZZXI, ZZXIX, ZXIXZ, XIXZZ \rangle,
  \end{equation}
  with the corresponding parity-check matrix given by
  \begin{equation}
    \label{eq: parity-check-bin}
    \mathbf{H} = 
    \begin{pNiceArray}{c:c}
      1 \, 0 \, 0 \, 1 \, 0 & 0 \, 1 \, 1 \, 0 \, 0 \\
      0 \, 1 \, 0 \, 0 \, 1 & 0 \, 0 \, 1 \, 1 \, 0 \\
      1 \, 0 \, 1 \, 0 \, 0 & 0 \, 0 \, 0 \, 1 \, 1 \\
      0 \, 1 \, 0 \, 1 \, 0 & 1 \, 0 \, 0 \, 0 \, 1 \\
    \end{pNiceArray}.
  \end{equation}
  Each stabilizer generator is represented as a row in the parity-check matrix, with the left~(right) half corresponding to $X$-type~($Z$-type) stabilizer.
  Once the parity-check matrix is set, one can construct the Tanner graph, as shown in Fig.~\ref{fig: tanner}. 
  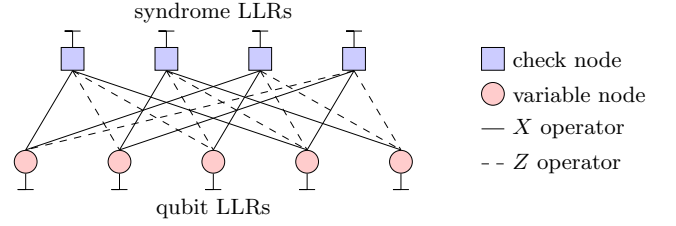
\begin{figure}[tb]
    \resizebox{.48\textwidth}{!}{\input{Figures/Tanner-graph.tex}}
    \caption{\label{fig: tanner}The Tanner graph of the $[[5, 1, 3]]$ perfect stabilizer code.}
  \end{figure}
  In the Tanner graph, each check node represents a row of the parity-check matrix, and each variable node represents a column.

  Assume the error rate for the quantum channel is $p$ and the error rate for the syndrome channel is $q$. 
  The initial soft information for the syndrome in the Tanner graph, used by the enhanced BP decoder, is given by the LLR $\pm \log \frac{1-q}{q}$.
  The sign of this LLR depends on whether the syndrome bit is $0$ or $1$. 
  Similarly, the initial qubit soft information is represented by a 3-tuple of LLRs:
  \begin{equation}
    \bigg(\log \frac{p_I}{p_X}, \log \frac{p_I}{p_Y}, \log \frac{p_I}{p_Z}\bigg)
  \end{equation}
  where $p_X$, $p_Y$ and $p_Z$ are the probabilities of the Pauli errors $X$, $Y$ and $Z$, respectively, and $p_I = 1 - p_X - p_Y - p_Z$.
  For a depolarizing channel with rate $p$, these probabilities are $p_X = p_Y = p_Z = p/3$ and $p_I = 1 - p$.
\end{example}
Once the Tanner graph is set up and the initial soft information is in place, the enhanced BP algorithm starts iterating over the graph. 
During each iteration, it updates the soft information for both the quantum information and the syndrome information. 
Messages are exchanged between the variable nodes and the check nodes, refining the LLRs based on the previous iteration's information and the Tanner graph's structure. 
To further improve performance, we incorporate normalized message passing in the enhanced BP algorithm. This technique uses a normalization factor to refine the pre-processing decoding, resulting in more accurate LLRs for both qubits and syndromes.
This iterative process continues until it either converges or reaches a predefined number of iterations, resulting in soft outputs that provide refined estimates for the error probabilities of the qubits and the syndromes. 

\subsection{OSD with refined LLRs}
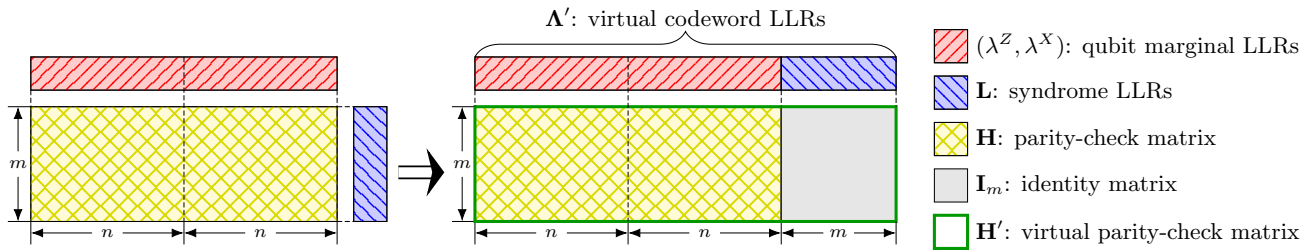
\begin{figure*}[tb]
  \resizebox{.975\textwidth}{!}{\input{Figures/concatenate.tex}}
  \caption{\label{fig: concatenate}Concatenation of qubits marginal LLRs and syndrome LLRs to form the LLRs of the new virtual codeword and simultaneous extension of the parity-check matrix to include the syndrome.}
\end{figure*}

The refined LLRs generated by the enhanced BP decoder will be used as input for the OSD algorithm, which will list and correct error patterns for both qubits and syndromes. 
To apply the OSD algorithm, we integrate the qubit and syndrome information to create a new virtual codeword. 
In order to do this, the binary parity-check matrix~\eqref{eq: parity-check-bin} must be extended to include the syndrome.
As shown in Fig.~\ref{fig: concatenate}, the new virtual codeword is formed by concatenating the quantum codeword and the syndrome, and the parity-check matrix is extended accordingly.
We continue Example~\ref{ex: codes} to illustrate the process of concatenating.
\begin{example}
  \label{ex: concatenate}
  Supppose that the soft output~(LLR) of the enhanced BP decoder in Example~\ref{ex: codes} is given as follows.
  \begin{equation}
    \mathbf{\Lambda} = 
    \begin{pNiceMatrix}[first-col]
      X &  2 &  0 &  4 &  0 &  4 \\
      Y &  1 & -3 &  4 &  0 &  1 \\
      Z &  2 & -3 &  4 &  0 &  1 \\
    \end{pNiceMatrix},
    \,
    \mathbf{L} = 
    \begin{pNiceMatrix}
      -4 & 1 & 3 & -5
    \end{pNiceMatrix}
  \end{equation}
  Here, the $j$-th column of $\mathbf{\Lambda}$~($1 \leq j \leq 5$) represents the LLR tuple for the $j$-th qubit, and the $i$-th element of $\mathbf{L}$~($1 \leq i \leq 4$) represents the LLR for the $i$-th syndrome.
  Then, the $X$- and $Z$-marginal LLRs for the $j$-th qubit are defined as
  \begin{equation}
    \begin{aligned}
      \lambda^X_j &= \log \frac{1 + \exp(-\Lambda_{j, Z})}{\exp(-\Lambda_{j, Y}) + \exp(-\Lambda_{j, X})},\\
      \lambda^Z_j &= \log \frac{1 + \exp(-\Lambda_{j, X})}{\exp(-\Lambda_{j, Y}) + \exp(-\Lambda_{j, Z})}. 
    \end{aligned}
  \end{equation}
  In our example, the $X$- and $Z$-marginal LLRs are calculated as
  \begin{equation}
    \begin{pNiceMatrix}
      \lambda^X \\
      \lambda^Z
    \end{pNiceMatrix}
    =
    \begin{pNiceMatrix}
      0.8 &  0 & 3.3 & 0 & 1.3 \\
      0.8 & -3 & 3.3 & 0 & 0.3
    \end{pNiceMatrix}.
  \end{equation}
  Finally, the LLR vector for the new virtual codeword is obtained by concatenating the $Z$- and $X$-marginal LLRs%
  \footnote{The $Z$-$X$ order of the marginal LLRs is opposite to the $X$-$Z$ order of the parity-check matrix~\eqref{eq: parity-check-bin} due to the symplectic inner product's alternating property.}
  with the syndrome LLRs, i.e.,
  \begin{equation}
    \begin{split}
      \label{eq: concatenate}
      &\mathbf{\Lambda}'  = 
      (\lambda^Z, \lambda^X \, | \, \mathbf{L}) = \\
      &\begin{pNiceArray}{c:c|c}
        \CodeBefore
        \columncolor{red!20}{1,2}
        \columncolor{blue!20}{3}
        \Body
      0.8, 0, 3.3, 0, 1.3 &
      0.8,-3, 3.3, 0, 0.3 &
      -4, 1, 3, -5
      \end{pNiceArray},
    \end{split}
  \end{equation}
  and the parity-check matrix is extended to include the syndrome, i.e.,
  \begin{equation}
    \mathbf{H}' = (\mathbf{H} \, | \, \mathbf{I}_4) =
    \begin{pNiceArray}{c:c|c}
      \CodeBefore
      \columncolor{yellow!20}{1,2}
      \columncolor{black!10}{3}
      \Body
      1 \, 0 \, 0 \, 1 \, 0 & 0 \, 1 \, 1 \, 0 \, 0 & 1 \, 0 \, 0 \, 0 \\
      0 \, 1 \, 0 \, 0 \, 1 & 0 \, 0 \, 1 \, 1 \, 0 & 0 \, 1 \, 0 \, 0 \\
      1 \, 0 \, 1 \, 0 \, 0 & 0 \, 0 \, 0 \, 1 \, 1 & 0 \, 0 \, 1 \, 0 \\
      0 \, 1 \, 0 \, 1 \, 0 & 1 \, 0 \, 0 \, 0 \, 1 & 0 \, 0 \, 0 \, 1 \\
    \end{pNiceArray}.
  \end{equation}
  The color shading in the matrices corresponds to Fig.~\ref{fig: concatenate}, providing an intuitive understanding.
\end{example}
Taking the extended parity-check matrix and the virtual codeword LLRs as input, the OSD algorithm will list $\ell_{\textrm{max}}$ virtual codewords and output the best one.
Note that the virtual codeword is not an actual codeword of the original quantum code; rather, it is a classical codeword in the finite field $GF(2)$ that encodes the estimated quantum and syndrome errors. 
In the next subsection, we will concentrate on extracting the quantum and syndrome errors from the codeword produced by the OSD algorithm.

\subsection{Extracting quantum/syndrome errors}
\label{subsec: extract}
The best virtual codeword $\bm{c} \in GF(2)^{2n+m}$ is selected from the list of candidates $\mathcal{L}$ generated by the OSD algorithm.
Here, the best virtual codeword is the one that minimizes the LLR metric $\Gamma(\cdot)$,
\begin{equation}
  \label{eq: quality}
  \bm{c} = \underset{\bm{v} \in \mathcal{L}}{\arg\min}\, \Gamma(\bm{v}),\, \Gamma(\bm{v}) = \sum_{i=1}^{2n+m} v_i \Lambda'_i.
\end{equation}
Next, we extract the quantum error pattern $\bm{e} \in \{I, X, Y, Z\}^{n}$ and the syndrome error pattern $\bm{s} \in GF(2)^{m}$ from the best virtual codeword $\bm{c}$.
The extraction process is illustrated in the following example~(Example~\ref{ex: extract}).

\begin{example}
  \label{ex: extract}
  The $i$-th Pauli operator of the quantum error pattern $\bm{e}$ and the $j$-th bit of the syndrome error pattern $\bm{s}$ are extracted as
  \begin{equation}
    \begin{aligned}
      e_i &= Z^{c_i}X^{c_{n + i}}, &1 \leq i \leq n,\\
      s_j &= c_{2n + j}, &1 \leq j \leq m.
    \end{aligned}
  \end{equation}
  In this context, $Z^0 = X^0 = I$ and $ZX = Y$.
  Assume the best virtual codeword $\bm{c} = 01000 01000 | 1101$ is chosen from the list $\mathcal{L}$ by the OSD in Example~\ref{ex: concatenate}.
  In this case, the extracted quantum error pattern is $\bm{e} = IYIII$, and the extracted syndrome error pattern is $\bm{s} = 1101$.
  After calculation, the syndrome of $\bm{e}$ coincides with $\bm{s}$, which is revealed in the following proposition~(Proposition~\ref{prop: syndrome}).
\end{example}

\begin{proposition}
  \label{prop: syndrome}
  Assume that the quantum channel error pattern $\bm{e}$ and the syndrome error pattern $\bm{s}$ are extracted from a valid virtual codeword $\bm{c}$. 
  Then, the syndrome of the Pauli error pattern $\bm{e}$ matches the estimated syndrome $\bm{s}$, i.e.,
  \begin{equation}
    \label{eq: syndrome}
    \mathbf{H} \odot \bm{e} = \bm{s}
  \end{equation}
  where $\mathbf{H}$ is the parity-check matrix of the quantum code, and $\odot$ denotes the symplectic inner product.
\end{proposition}
\begin{proof}
  The proof is straightforward by substituting the binary form of $\bm{e}$ and $\bm{s}$.
  \begin{align}
    \mathbf{H} \odot \bm{e}
    &= (\mathbf{H}^X \odot \bm{e}^Z) \oplus (\mathbf{H}^Z \odot \bm{e}^X) \\
    &= \mathbf{H} (\bm{c}_L)^{\mathrm{T}} \\
    &= \mathbf{H}' \bm{c}^{\mathrm{T}} \oplus \mathbf{I}_m(\bm{c}_R)^{\mathrm{T}}\\
    &= \bm{0} \oplus \bm{s} = \bm{s}.
  \end{align}
  Here, $\oplus$ denotes the bitwise XOR operation, and $\bm{c}_L$~(the first $2n$ bits) and $\bm{c}_R$~(the last $m$ bits) represent segments of the virtual codeword $\bm{c}$.

\end{proof}

\subsection{Parameters and details}
\label{subsec: tuning}
We have now introduced the extended BP-OSD algorithm, which includes several parameters that can impact both decoding performance and computational complexity. 
Some of the most crucial parameters among them are:
\begin{enumerate}
  \item The maximum iteration number $T_{\textrm{max}}$ and normalization factor $\alpha$ for the enhanced BP algorithm;
  \item The syndrome weight $\beta$ to adjust LLRs fed into the OSD algorithm;
  \item The maximum list size $\ell_{\textrm{max}}$ for the OSD algorithm.
\end{enumerate}

The BP normalization factor $\alpha$ is used to normalize messages from variable nodes to check nodes in the Tanner graph. 
Specifically, we use the normalized min-sum~(NMS) algorithm for BP~\cite{poulin2008iterative}, where the message from a variable node to a check node is multiplied by the factor $\alpha$, while the message from a check node to a variable node remains unchanged.
In the proposed decoding algorithm, we use two normalization factors, $\alpha_1$ and $\alpha_2$, for the BP algorithm.
\begin{itemize}
  \item First stage (NMS with $\alpha_1$):
  \begin{itemize}
    \item Run the NMS algorithm with $\alpha_1$.
    \item Check if the estimated syndrome and Pauli error pattern are consistent~(see~\eqref{eq: syndrome}).
    \item If consistent, output the estimated Pauli error pattern and skip the second stage.
  \end{itemize}
  \item Second stage (NMS with $\alpha_2$):
  \begin{itemize}
    \item Run the NMS algorithm again but with $\alpha_2$.
    \item Continue with the remaining tasks as described previously (see Fig.~\ref{fig: framework}).
  \end{itemize}
\end{itemize}

The syndrome weight $\beta$ is used to balance the importance of quantum information and syndrome information before the soft information from BP is fed into the OSD algorithm. 
The balance is achieved by multiplying the syndrome LLRs by the factor $\beta$, i.e., \eqref{eq: concatenate} is modified as 
\begin{equation}
  \mathbf{\Lambda}'  = (\lambda^Z, \lambda^X \, | \, \beta \mathbf{L}).
\end{equation}

\subsection{Complexity analysis}
\label{subsec: complexity}
The complexity of the extended BP-OSD algorithm is primarily influenced by the two components it employs: the BP part and the OSD part.
The BP algorithm has an average complexity%
\footnote{The BP algorithm generally has a complexity of $O(n w T_{\textrm{avg}})$, where $w$ is the average degree in the Tanner graph. For surface codes, $w$ is at most $4$.}
of $O(nT_{\textrm{avg}})$, where $n$ is the number of qubits and $T_{\textrm{avg}}$ is the average number of iterations.
Specially, we use $T_{\textrm{1,avg}}$~(resp. $T_{\textrm{2,avg}}$) to denote the average number of iterations in the first~(resp. second) stage of the BP algorithm.
The OSD algorithm, on the other hand, has an average complexity of $O(nm\ell_{\textrm{avg}})$, where $\ell_{\textrm{avg}}$ is the average list size~\cite{liang2024random}. 
If we denote the probability of executing the second stage as $\eta$, the overall average complexity of the extended BP-OSD algorithm is:
\begin{align}
  \underbrace{O(n T_{\textrm{1,avg}}) }_{\textrm{First stage}} + 
  \eta \cdot \underbrace{O(n (T_{\textrm{2,avg}} + m\ell_{\textrm{avg}}))}_{\textrm{Second stage}}.
\end{align}
When physical error rates are low, the BP algorithm converges quickly, which implies $\eta$ approaches $0$, i.e., the second stage is rarely executed, resulting in a close complexity to the BP algorithm.
This is numerically verified in Sec.~\ref{sec: simulation}.

\section{Simulation}
\label{sec: simulation}
In this section, we present the numerical results of the proposed BP-OSD algorithm.
For comparison, we also include the results of the MWPM decoder, BP decoder without normalization, NMS decoder and the original BP-OSD decoder.
\subsection{Simulation setup}
We evaluate the performance of the proposed BP-OSD algorithm on the $[[2d^2 - 2d + 1, 1, d]]$ surface code.
The setup and simulated algorithms are summarized in Table~\ref{tab: parameters} and~\ref{tab: algorithms} respectively.
\begin{table}[tb]
  \caption{\label{tab: parameters}Simulation parameters.}
  \begin{tabular}{l|cc}
      \hline
      \hline
      {Description} & {Symbol} & {Values} \\
      \hline
      Distance of the surface code & $d$ & $5$\\
      BP maximum iteration number & $T_{\textrm{max}}$ & $32$\\
      BP normalization factors & $\alpha_1, \alpha_2$ & ${5/8, 1}$\\
      Syndrome LLRs weight & $\beta$ & $7.5$\\
      OSD maximum list size & $\ell_{\textrm{max}}$ & $2^{10}$\\
      Depolarizing rate & $p$ & $10^{-4}$ to $1$\\
      Syndrome bit-flip rate & $q$ & $10^{-5}$ to $10^{-3}$\\
      \hline
      \hline
  \end{tabular}
\end{table}
\begin{table}[tb]
  \caption{\label{tab: algorithms}Simulated decoding algorithms.}
  \begin{tabular}{l|ccc}
      \hline
      \hline
      {Algorithm} & {Syndrome}\footnotemark[1] & {Preprocess} & {Postprocess} \\
      \hline
      MWPM & $\hat{q}=0$ & n/a & MWPM~\cite{fowler2013minimum}\\
      original BP & $\hat{q}=0$ & BP~\cite{ching2021log} & hard decision\\
      original BP-OSD & $\hat{q}=0$ & BP & OSD~\cite{panteleev2021degenerate}\\
      enhanced BP & $\hat{q}=q$ & enhanced BP & hard decision \\
      extended BP-OSD & $\hat{q}=q$ & enhanced BP & extended OSD\\
      \hline
      \hline
  \end{tabular}
  \footnotetext[1]{Here, $\hat{q}=0$ ignores the possibly exsiting erroneous mesurments and assumes an error-free syndrome, while $\hat{q}=q$ considers the syndrome bit-flip rate of $q$.}
\end{table}

\subsection{Numerical results}
\label{subsec: results}
\begin{figure}[tb]
  \resizebox{.48\textwidth}{!}{\input{Figures/plot-SameSyndrome.tex}}
  \caption{\label{fig: plot-samesyndrome}Syndrome error rate of the $[[41, 1, 5]]$ surface code by different decoders. The syndrome bit-flip rate $q=10^{-5}$ is fixed.}
\end{figure}
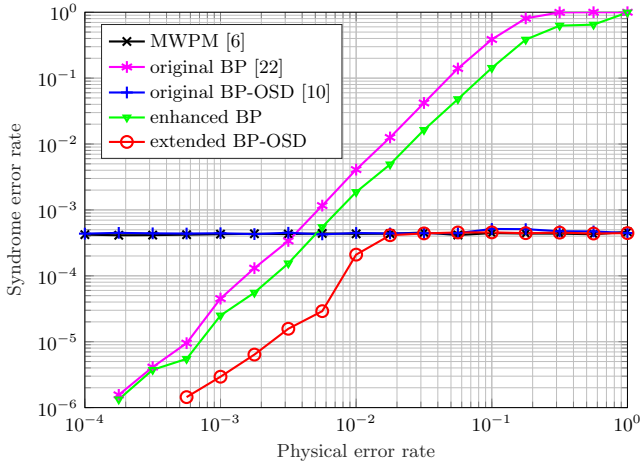
We first investigate the syndrome error rate of the $[[41, 1, 5]]$ surface code using different decoders. 
Fig.~\ref{fig: plot-samesyndrome} shows the syndrome error rate, defined as the probability that the estimated syndrome is incorrect, under various quantum channel error rates.
As illustrated, the syndrome error rate of the MWPM and the original BP-OSD decoder remain constant regardless of the quantum channel error rate, since they assume the syndrome is always correct. 
The performance of the these two decoder can be estimated by $1 - (1 - q)^{m}$, indicating that it will frequently fail when the syndrome error rate is high or the length of the code is long.
The original BP decoder performs better than these two decoders at low quantum channel error rates, but its performance deteriorates as the error rate increases, eventually becoming worse than the MWPM decoder at high error rates.
This is because the BP decoder can effectively use qubit LLRs at low physical error rates, but it struggles to correct syndromes with errors at higher physical error rates.
The enhanced BP decoder is the BP decoder with consideration of syndrome LLRs, which is slightly better than the original BP decoder.
The extended BP-OSD algorithm outperforms all of the other decoders. 
More precisely, the extended BP-OSD algorithm achieves a syndrome error rate close to that of the original BP-OSD and the MWPM decoders at high quantum channel error rates and surpasses them at low error rates.
Moreover, the extended BP-OSD algorithm is approximately ten times more effective than the other two BP decoders in terms of syndrome error rate.
We can conclude from these results that the extended BP-OSD algorithm effectively corrects syndromes with errors, which is crucial for decoding surface codes with erroneous measurements.

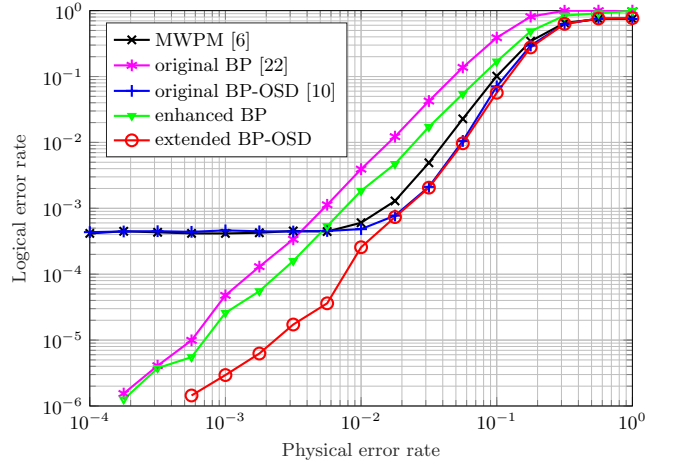
\begin{figure}[tb]
  \resizebox{.48\textwidth}{!}{\input{Figures/plot-EqNoise.tex}}
  \caption{\label{fig: plot-eqnoise}Logical error rate of the $[[41, 1, 5]]$ surface code by different decoders. The syndrome bit-flip rate $q=10^{-5}$ is fixed.}
\end{figure}

We then evaluate the logical error rate of the $[[41, 1, 5]]$ surface code using different decoders.
Fig.~\ref{fig: plot-eqnoise} shows the logical error rate, defined as the probability that the estimated Pauli error is not equivalent to the actual error under the stabilizer group.
We observe that the proposed BP-OSD algorithm outperforms other decoders in terms of logical error rate. 
This superior performance is primarily due to the significant improvements made by the OSD algorithm in post-processing. Additionally, the extended BP-OSD algorithm's ability to correct errors in the syndrome contributes to its advantage over the MWPM decoder and the original BP-OSD decoder, especially at low physical error rates.

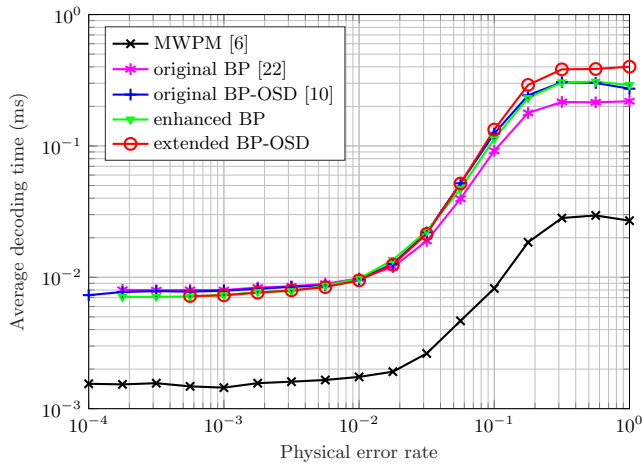
\begin{figure}[tb]
  \resizebox{.48\textwidth}{!}{\input{Figures/plot-EqNoise-ms.tex}}
  \caption{\label{fig: plot-eqnoise-ms} Average decoding time of the $[[41, 1, 5]]$ surface code by different decoders. The syndrome bit-flip rate $q=10^{-5}$ is fixed.}
\end{figure}
To further evaluate the efficiency of the proposed BP-OSD algorithm, we analyze its average decoding time compared to other decoders, as shown in Fig.~\ref{fig: plot-eqnoise-ms}.
The figure shows that at low physical error rates, the extended BP-OSD algorithm has a similar or even shorter decoding time compared to the original BP decoder. 
This is because the BP-OSD algorithm often skips the second stage, leading to a lower average decoding time, as discussed in Sec.~\ref{subsec: complexity}.

Finally, we compare the logical error rate at different syndrome bit-flip rates. 
As shown in Table~\ref{tab: logical-error-rate}, the BP-OSD algorithm achieves the lowest logical error rate compared to other decoders, especially at low syndrome bit-flip rates. 
This demonstrates its superior performance in correcting both quantum channel errors and syndrome errors.
\begin{table}[tb]
  \caption{\label{tab: logical-error-rate}Logical error rate with different syndrome bit-flip rates $q$, where the physical error rate is fixed at $p = 10^{-2.5}$.}
  \begin{tabular}{l|ccccc|c}
      \hline
      \hline
      $q$ & MWPM & \makecell{orignial\\BP} & \makecell{orignial\\BP-OSD} & \makecell{enhanced\\BP} & \makecell{extended\\BP-OSD} &  Scale \\
      \hline
      $10^{-5}$   & $33.9$ & $21.8$ & $15.6$ & $15.8$ & $\mathbf{1.68}$ & $\times 10^{-5}$ \\
      $10^{-4.5}$ & $35.6$ & $22.3$ & $4.60$ & $18.0$ & $\mathbf{3.48}$ & $\times 10^{-5}$ \\
      $10^{-4}$   & $4.14$ & $2.42$ & $1.12$ & $2.26$ & $\mathbf{0.93}$ & $\times 10^{-4}$ \\
      $10^{-3.5}$ & $6.76$ & $4.53$ & $3.71$ & $4.94$ & $\mathbf{3.44}$ & $\times 10^{-4}$ \\
      $10^{-3}$   & $1.16$ & $1.04$ & $1.18$ & $1.08$ & $\mathbf{1.01}$ & $\times 10^{-3}$ \\

      \hline
      \hline
  \end{tabular}
\end{table}

\section{Conclusion and discussion}
\label{sec: conclusion}
In this paper, we have proposed an extended BP-OSD algorithm for decoding surface codes with erroneous syndromes. 
Our method integrates syndrome soft information in the BP decoder using normalized message passing for pre-processing and incorporates virtual codeword with extended matrix into the OSD algorithm for post-processing.
The numerical results demonstrate that our proposed algorithm effectively recovers erroneous syndromes and significantly enhances the decoding performance of surface codes compared to the MWPM, original BP and original BP-OSD algorithms.
The proposed BP-OSD algorithm achieves a lower syndrome error rate and logical error rate, making it more effective in correcting errors in quantum channels and syndromes. 
Notably, the proposed algorithm lists quantum bits and syndromes simultaneously without requiring extra measurements and with low additional computational overhead.
This feature sets it apart from many existing decoding algorithms and is crucial for practical quantum error correction, where syndromes are often erroneous due to noise in the measurement process.
Our findings suggest that the extended BP-OSD algorithm can be a valuable tool for enhancing the reliability of quantum error correction in practical quantum computing systems. 
Future work could explore further optimizations of the extended BP-OSD algorithm, such as reducing its computational complexity and latency, and extending its applicability to other types of quantum error-correcting codes.

\bibliography{bibliography}

\end{document}

%% file: Figures/surface-stabilizer.tex
\begin{tikzpicture}
  \def\colnum{6}
  \def\rownum{4}
  \foreach \x in {2,...,\colnum} 
  {
    \draw[line width=1.5pt,line cap=round,color=gray] (\x,-.5) -- (\x,\rownum-.5);
  }
  
  \foreach \y in {0,...,\rownum} 
  {
    \draw[line width=1.5pt,line cap=round,color=gray] (2,\y-.5) -- (\colnum,\y-.5);
  }
  
  \draw[line width=4.5pt,line cap=round,color=white] (3, \rownum-.65) -- (3, \rownum-2.35);
  \draw[line width=4.5pt,line cap=round,color=white] (2.15, \rownum-1.5) -- (3.85, \rownum-1.5);
  
  \draw[line width=3pt,line cap=round,color=red!75!black] (3, \rownum-.65) -- (3, \rownum-1.275);
  \draw[line width=3pt,line cap=round,color=red!75!black] (3, \rownum-1.725) -- (3, \rownum-2.35);
  \draw[line width=3pt,line cap=round,color=red!75!black] (2.15, \rownum-1.5) -- (2.725, \rownum-1.5);
  \draw[line width=3pt,line cap=round,color=red!75!black] (3.225, \rownum-1.5) -- (3.85, \rownum-1.5);

  \draw[line width=4.5pt,line cap=round,color=white] (5.85-1,\rownum-1.5-2) -- (5.15-1,\rownum-1.5-2);
  \draw[line width=4.5pt,line cap=round,color=white] (5-1,\rownum-1.35-2) -- (5-1,\rownum-.65-2);
  \draw[line width=4.5pt,line cap=round,color=white] (5.15-1,\rownum-.5-2) -- (5.85-1,\rownum-.5-2);
  \draw[line width=4.5pt,line cap=round,color=white] (6-1,\rownum-.65-2) -- (6-1,\rownum-1.35-2);
  \draw[line width=3pt,line cap=round,color=blue!75!black] (5.85-1,\rownum-1.5-2) -- (5.15-1,\rownum-1.5-2);
  \draw[line width=3pt,line cap=round,color=blue!75!black] (5-1,\rownum-1.35-2) -- (5-1,\rownum-.65-2);
  \draw[line width=3pt,line cap=round,color=blue!75!black] (5.15-1,\rownum-.5-2) -- (5.85-1,\rownum-.5-2);
  \draw[line width=3pt,line cap=round,color=blue!75!black] (6-1,\rownum-.65-2) -- (6-1,\rownum-1.35-2);

  \node at (3, \rownum-1.5) {$v$};
  \node at (3.5, \rownum-1) {\textcolor{red!75!black}{$Z_v$}};
  \node at (5.5-1, \rownum-1-2) {$f$};
  \node at (5.5-1, \rownum-2-2) {\textcolor{blue!75!black}{$X_f$}};
\end{tikzpicture}

%% file: Figures/system-model.tex
\begin{tikzpicture}[
  block/.style={
  rectangle,
  draw,
  align=center,
  minimum width=1cm,
  minimum height=1.3cm,
}]
\node (depolar) [block] {depolarizing\\channel};
\node (dot) [circle,fill=black,minimum size=.3em,inner sep=0pt,right=3.77em of depolar] {};
\node (stabilizer) [block,below=2.4em of dot] {ideal\\measurement};
\node (bsc) [block,right=4.4em of stabilizer] {bit-flip\\channel};
\draw[dash pattern=on 4pt off 2pt, line width=.6pt] (3.3em,-2.2em) rectangle +(15.65em,-4.75em);
\node[below] at (11.125em,-6.75em) {stabilizer measurement with errors};
\draw ([xshift=-3em]depolar.west) -- (depolar) node[midway, align=center] {sent\\qubits};
\draw (dot.center) -- (stabilizer);
\draw[double distance=.15em] (stabilizer) -- (bsc) node[midway, align=center] {ideal\\syndrome};
\draw[double distance=.15em] (bsc) -- ([xshift=5em]bsc.east) node[pos=.55, align=center] {corrupted\\syndrome};
\draw (depolar) -- ([xshift=20.6em]depolar.east) node[pos=0.9, align=center] {received\\qubits};
\end{tikzpicture}

%% file: Figures/depolarizing-channel.tex
\begin{tikzpicture}
  \node (psi0) {$|\psi\rangle$};
  \node (psi1) [rectangle,minimum width=3em,right=10em of psi0] {$|\psi\rangle$};
  \node (psi2) [rectangle,minimum width=3em,below=1em of psi1] {$X|\psi\rangle$};
  \node (psi3) [rectangle,minimum width=3em,below=1em of psi2] {$Y|\psi\rangle$};
  \node (psi4) [rectangle,minimum width=3em,below=1em of psi3] {$Z|\psi\rangle$};
  
  \draw[{-{Latex}}] (psi0.east) -- (psi1.west);
  \draw[{-{Latex}}] (psi0.east) -- (psi2.west);
  \draw[{-{Latex}}] (psi0.east) -- (psi3.west);
  \draw[{-{Latex}}] (psi0.east) -- (psi4.west);
  \node at(3,-.2 ) {$1 - p$};
  \node at(3,-.94) {$p / 3$};
  \node at(3,-1.71) {$p / 3$};
  \node at(3,-2.49) {$p / 3$};
  \end{tikzpicture}

%% file: Figures/bit-flip-channel.tex
\begin{tikzpicture}
  \node (0l) {$0$};
  \node (1l) [below=of 0l] {$1$};
  \node (0r) [right=10em of 0l] {$0$};
  \node (1r) [below=of 0r] {$1$};

  \draw[{-{Latex}}] (0l.east) -- (0r.west)node[midway, above] {$1 - q$};
  \draw[{-{Latex}}] (0l.east) -- (1r.west)node[pos=.15, below] {$q$};
  \draw[{-{Latex}}] (1l.east) -- (1r.west)node[midway, below] {$1 - q$};
  \draw[{-{Latex}}] (1l.east) -- (0r.west)node[pos=.15, above] {$q$};
\end{tikzpicture}

%% file: Figures/decoder-goal.tex
\begin{tikzpicture}[
  block/.style={
  rectangle,
  draw,
  align=center,
  minimum width=1.7cm,
  minimum height=1.3cm,
}]
\node (bplc) [block] {proposed\\method};
\node (ecorr) [block] at ([yshift=2.5em, xshift=6em]bplc.north east) {recovery\\operations};

\draw[double distance=.15em] ([xshift=-5em]bplc.west) -- (bplc) node[midway, align=center] {corrupted\\syndrome};
\draw[double distance=.15em] (bplc) -| (ecorr) node[pos=0.245, align=center] {estimated\\Pauli errors};
\draw ([xshift=-13.65em]ecorr.west) -- (ecorr) node[pos=0.15, align=center] {received\\qubits};
\draw (ecorr) -- ([xshift=5em]ecorr.east) node[midway, align=center] {corrected\\qubits};
\end{tikzpicture}

%% file: Figures/framework.tex
\begin{tikzpicture}[
  block/.style={
  rectangle,
  draw,
  align=center,
  minimum width=5em,
  minimum height=2.5em,
},
XTIME/.style={draw,circle,minimum width=1.5em,minimum height=1.5em,append after command={
[shorten >=\pgflinewidth, shorten <=\pgflinewidth,]
(\tikzlastnode.north east) edge (\tikzlastnode.south west)
(\tikzlastnode.south east) edge (\tikzlastnode.north west)
}
}]

\node (nms) [block] {BP($T_{\mathrm{max}},\alpha$)};
\node(marg) [block,right=3.5em of nms] {marginalize};
\node(conca) [block,right=7em of marg] {concatenation};
\node(time) [XTIME,below=2em of marg,xshift=-1.25em] {};
\node(osd) [block,below=7em of conca] {OSD($\ell_{\mathrm{max}}$)};
\node(extra) [block,below=7em of marg,xshift=-4em] {extraction};

\draw[-{Latex}] (nms)++(0,3em) -- (nms) node[pos=0,above, align=center] {corrupted\\syndrome};
\draw[-{Latex}] (nms) -- (marg) node[midway, align=center] {qubit\\LLRs};
\draw[-{Latex}] (marg.east) ++(0,.4em) -- ++(7em,0) node[midway,above] {\footnotesize{$X$-margin LLRs}};
\draw[-{Latex}] (marg.east) ++(0,-.4em) -- ++(7em,0) node[midway,below] {\footnotesize{$Z$-margin LLRs}};
\draw[-{Latex}] (nms) |- (time) node[pos=.75,above] {syndrome LLRs};
\draw[{Latex}-] (conca.south)+(-1.5em,0) |- (time) node[pos=.75,above] {weighted syndrome LLRs}; 
\draw[-{Latex}] (time)+(0,-2.2em) -- (time) node[pos=.35,right] {$\beta$};
\draw[-{Latex}] (conca.south)++(2em, 0) -- ++(0,-7em) node[pos=.8,left] {virtual codeword LLRs};
\draw[-{Latex}] (osd) -- (extra) node[midway,above] {virtual codeword};
\draw[-{Latex}] (osd)+(0,-3em) -- (osd) node[pos=.4,left] {virtual parity-check matrix};
\draw[{Latex}-] (nms)++(0,-12.5em) |- (extra) node[pos=0,below, align=center] {estimated\\Pauli error};
\end{tikzpicture}

%% file: Figures/Tanner-graph.tex
\begin{tikzpicture}[
  var/.style={circle,fill=red!20!white,draw,inner sep=0pt,minimum size=1em,append after command={\pgfextra{\draw (\tikzlastnode.south) -- ++(0,-.7em);\draw (\tikzlastnode.south) ++(0,-0.7em) -- ++(-0.35em,0) -- ++(0.7em,0);}}},
  check/.style={rectangle,fill=blue!20!white,draw,inner sep=0pt,minimum size=1em,append after command={\pgfextra{\draw (\tikzlastnode.north) -- ++(0,.7em);\draw (\tikzlastnode.north) ++(0,0.7em) -- ++(-0.35em,0) -- ++(0.7em,0);}}}]
  
  \node (c0) [check] {};
  \node (c1) [check, right=of c0] {};
  \node (c2) [check, right=of c1] {};
  \node (c3) [check, right=of c2] {};
  \node (v1) at ($(c0)!0.5!(c1)$) [var, below=4em] {};
  \node (v0) [var, left=of v1] {};
  \node (v2) [var, right=of v1] {};
  \node (v3) [var, right=of v2] {};
  \node (v4) [var, right=of v3] {};
  
  \draw         (c0.south) -- (v0.north);
  \draw[dashed] (c0.south) -- (v1.north);
  \draw[dashed] (c0.south) -- (v2.north);
  \draw         (c0.south) -- (v3.north);
  
  \draw         (c1.south) -- (v4.north);
  \draw         (c1.south) -- (v1.north);
  \draw[dashed] (c1.south) -- (v2.north);
  \draw[dashed] (c1.south) -- (v3.north);
  
  \draw         (c2.south) -- (v0.north);
  \draw[dashed] (c2.south) -- (v4.north);
  \draw         (c2.south) -- (v2.north);
  \draw[dashed] (c2.south) -- (v3.north);
  
  \draw[dashed] (c3.south) -- (v0.north);
  \draw         (c3.south) -- (v1.north);
  \draw[dashed] (c3.south) -- (v4.north);
  \draw         (c3.south) -- (v3.north);
  
  \node (top) [above=5.25em of v2] {syndrome LLRs};
  \node (down) [below=.75em of v2] {qubit LLRs};
  
  \begin{scope}[shift={(6,0)}]
    \node at (0,-.5) [circle,fill=red!20!white,draw,minimum size=1em, label=right:variable node] {};
    \node at (0,0) [rectangle,fill=blue!20!white,draw,minimum size=1em, label=right:check node] {};
    \draw (-.5em,-1) -- (.5em,-1) node[right] {$X$ operator};
    \draw[dashed] (-.5em,-1.5) -- (.5em,-1.5) node[right] {$Z$ operator};
  \end{scope}
\end{tikzpicture}

%% file: Figures/concatenate.tex
\begin{tikzpicture}
  \def\h{13pt}
  \def\r{45pt}
  \def\n{60pt}
  
  \draw[fill=red!20!white] (0,0) rectangle +(2*\n,\h);
  \draw[fill opacity=1,pattern={Lines[angle=45,distance={3.5pt},line width=.5pt]},pattern color=red] (0,0) rectangle +(2*\n,\h);
  
  \draw[fill=yellow!20!white] (0,-.5*\h) rectangle +(2*\n,-\r);
  \draw[fill opacity=1,pattern=hatch,pattern color=yellow!85!black,hatch size=7pt,hatch line width=.7pt] (0,-.5*\h) rectangle +(2*\n,-\r);
  
  \draw[dash=on 2pt off 1pt phase  0pt] (0,0) -- (0,-.5*\h);
  \draw[dash=on 2pt off 1pt phase  0pt] (2*\n,0) -- (2*\n,-.5*\h);
  \draw[dash=on 2pt off 1pt phase  0pt] (\n,\h) -- (\n,-\r-.5*\h);
  
  \draw (0*\h,-.5*\h) -- (-9+0*\h,-.5*\h);
  \draw (0*\h,-.5*\h-\r) -- (-9+0*\h,-.5*\h-\r);
  \draw[{-{Latex}}] (-5+0*\h, -.5*\h-.5*\r+5) -- (-5+0*\h, -.5*\h);
  \draw[{-{Latex}}] (-5+0*\h, -.5*\h-.5*\r-5) -- (-5+0*\h, -.5*\h-\r);
  \node at (0*\h-5,-.5*\h-.5*\r) {\scriptsize$m$};
  
  \draw (0,-.5*\h-\r) -- (0,-.5*\h-\r-9);
  \draw (\n,-.5*\h-\r) -- (\n,-.5*\h-\r-9);
  \draw (2*\n,-.5*\h-\r) -- (2*\n,-.5*\h-\r-9);
  \draw[{-{Latex}}] (.5*\n-6, -.5*\h-\r-4.5) -- (0, -.5*\h-\r-4.5);
  \draw[{-{Latex}}] (.5*\n+6, -.5*\h-\r-4.5) -- (\n, -.5*\h-\r-4.5);
  \draw[{-{Latex}}] (1.5*\n-6, -.5*\h-\r-4.5) -- (\n, -.5*\h-\r-4.5);
  \draw[{-{Latex}}] (1.5*\n+6, -.5*\h-\r-4.5) -- (2*\n, -.5*\h-\r-4.5);
  \node at (.5*\n, -.5*\h-\r-4.5) {\scriptsize$n$};
  \node at (1.5*\n, -.5*\h-\r-4.5) {\scriptsize$n$};
  
  \draw[dash=on 2pt off 1pt phase  0pt] (2*\n,-.5*\h) -- (2*\n+.5*\h,-.5*\h);
  \draw[dash=on 2pt off 1pt phase  0pt] (2*\n,-.5*\h-\r) -- (2*\n+.5*\h,-.5*\h-\r);
  
  \draw[fill=blue!20!white] (.5*\h+2*\n,-.5*\h) rectangle +(\h,-\r);
  \draw[fill opacity=1,pattern={Lines[angle=135,distance={3.5pt},line width=.5pt]},pattern color=blue] (.5*\h+2*\n,-.5*\h) rectangle +(\h,-\r);
  
  \begin{scope}[xshift=2.9*\n]
    \draw[fill=red!20!white] (0,0) rectangle +(2*\n,\h);
    \draw[fill opacity=1,pattern={Lines[angle=45,distance={3.5pt},line width=.5pt]},pattern color=red] (0,0) rectangle +(2*\n,\h);
    
    \draw[fill=yellow!20!white] (0,-.5*\h) rectangle +(2*\n,-\r);
    \draw[fill opacity=1,pattern=hatch,pattern color=yellow!85!black,hatch size=7pt,hatch line width=.7pt] (0,-.5*\h) rectangle +(2*\n,-\r);
    
    \draw (0*\h,-.5*\h) -- (-9+0*\h,-.5*\h);
    \draw (0*\h,-.5*\h-\r) -- (-9+0*\h,-.5*\h-\r);
    \draw[{-{Latex}}] (-5+0*\h, -.5*\h-.5*\r+5) -- (-5+0*\h, -.5*\h);
    \draw[{-{Latex}}] (-5+0*\h, -.5*\h-.5*\r-5) -- (-5+0*\h, -.5*\h-\r);
    \node at (0*\h-5,-.5*\h-.5*\r) {\scriptsize$m$};
  
    \draw (0,-.5*\h-\r) -- (0,-.5*\h-\r-9);
    \draw (\n,-.5*\h-\r) -- (\n,-.5*\h-\r-9);
    \draw (2*\n,-.5*\h-\r) -- (2*\n,-.5*\h-\r-9);
    \draw[{-{Latex}}] (.5*\n-6, -.5*\h-\r-4.5) -- (0, -.5*\h-\r-4.5);
    \draw[{-{Latex}}] (.5*\n+6, -.5*\h-\r-4.5) -- (\n, -.5*\h-\r-4.5);
    \draw[{-{Latex}}] (1.5*\n-6, -.5*\h-\r-4.5) -- (\n, -.5*\h-\r-4.5);
    \draw[{-{Latex}}] (1.5*\n+6, -.5*\h-\r-4.5) -- (2*\n, -.5*\h-\r-4.5);
    \node at (.5*\n, -.5*\h-\r-4.5) {\scriptsize$n$};
    \node at (1.5*\n, -.5*\h-\r-4.5) {\scriptsize$n$};
  
    \draw[fill=blue!20!white] (2*\n,0) rectangle +(\r, \h);
    \draw[fill opacity=1,pattern={Lines[angle=135,distance={3.5pt},line width=.5pt]},pattern color=blue] (2*\n,0) rectangle +(\r, \h);
  
    \draw[fill=black!10] (2*\n,-.5*\h) rectangle +(\r,-\r);
    
    \draw[fill opacity=1,green!60!black,very thick] (0,-.5*\h) rectangle +(2*\n+\r,-\r);
    
    \draw[dash=on 2pt off 1pt phase  0pt] (0,0) -- (0,-.5*\h);
    \draw[dash=on 2pt off 1pt phase  0pt] (2*\n,0) -- (2*\n,-.5*\h);
    \draw[dash=on 2pt off 1pt phase  0pt] (2*\n+\r,0) -- (2*\n+\r,-.5*\h);
    \draw[dash=on 2pt off 1pt phase  0pt] (\n,\h) -- (\n,-\r-.5*\h);
  
    \draw (2*\n+\r,-.5*\h-\r) -- (2*\n+\r,-.5*\h-\r-9);
    \draw[{-{Latex}}] (2*\n+.5*\r-6, -.5*\h-\r-4.5) -- (2*\n, -.5*\h-\r-4.5);
    \draw[{-{Latex}}] (2*\n+.5*\r+6, -.5*\h-\r-4.5) -- (2*\n+\r, -.5*\h-\r-4.5);
    \node at (2*\n+.5*\r, -.5*\h-\r-4.5) {\scriptsize$m$};
  
    \draw [decorate, decoration={brace, amplitude=10pt}] (0,\h) -- (2*\n+\r,\h) node [black,midway,yshift=.55cm] {$\mathbf{\Lambda}'$: virtual codeword LLRs };
  \end{scope}
  
  \begin{scope}[xshift=5.9*\n,yshift=.825*\h]
    \draw[fill=red!20!white] (0,0) rectangle +(\h,\h);
    \draw[fill opacity=1,pattern={Lines[angle=45,distance={3.5pt},line width=.5pt]},pattern color=red]  (0,0) rectangle +(\h,\h);
    \node[anchor=west] at (\h,.5*\h) {$(\lambda^Z, \lambda^X)$: qubit marginal LLRs};
  
    \draw[fill=blue!20!white] (0,0-1.4*\h) rectangle +(\h,\h);
    \draw[fill opacity=1,pattern={Lines[angle=135,distance={3.5pt},line width=.5pt]},pattern color=blue]  (0,0-1.4*\h) rectangle +(\h,\h);
    \node[anchor=west] at (\h,.5*\h-1.4*\h) {$\mathbf{L}$: syndrome LLRs};
  
    \draw[fill=yellow!20!white] (0,0-2*1.4*\h) rectangle +(\h,\h);
    \draw[fill opacity=1,pattern=hatch,pattern color=yellow!85!black,hatch size=7pt,hatch line width=.7pt] (0,0-2*1.4*\h) rectangle +(\h,\h);
    \node[anchor=west] at (\h,.5*\h-2*1.4*\h) {$\mathbf{H}$: parity-check matrix };
  
    \draw[fill=black!10] (0,0-3*1.4*\h) rectangle +(\h,\h);
    \node[anchor=west] at (\h,.5*\h-3*1.4*\h) {$\mathbf{I}_{m}$: identity matrix };
  
    \draw[green!60!black,very thick] (0,0-4*1.4*\h) rectangle +(\h,\h);
    \node[anchor=west] at (\h,.5*\h-4*1.4*\h) {$\mathbf{H}'$: virtual parity-check matrix };
  \end{scope}
  
  \draw [line width=.7pt, double distance=4pt,arrows = {-Stealth[width'=5pt 2, length=7pt]}] (2.4*\n,-.5*\r-.75*\h) -- (2.7*\n,-.5*\r-.75*\h);
\end{tikzpicture}

%% file: Figures/plot-SameSyndrome.tex
%
%
\definecolor{mycolor1}{rgb}{1.00000,0.00000,1.00000}%
\begin{tikzpicture}[%
thick,scale=1.00, every node/.style={scale=1.00}
]

\begin{axis}[%
width=9.066666699999999cm,
height=6.6115258cm,
at={(0cm,0cm)},
scale only axis,
xmode=log,
xmin=0.0001,
xmax=1,
xminorticks=true,
xlabel style={font=\color{white!15!black}},
xlabel={Physical error rate},
ymode=log,
ymin=1e-06,
ymax=1,
yminorticks=true,
ylabel style={font=\color{white!15!black}},
ylabel={Syndrome error rate},
axis background/.style={fill=white},
xmajorgrids,
xminorgrids,
ymajorgrids,
yminorgrids,
legend style={at={(0.03,0.97)}, anchor=north west, legend cell align=left, align=left, draw=white!15!black}
]
\addplot [color=black, line width=1.0pt, mark size=2.8pt, mark=x, mark options={solid, black}]
  table[row sep=crcr]{%
1	0.000464278418482\\
0.562341325190349	0.000427409137397\\
0.316227766016838	0.000437015186278\\
0.177827941003892	0.000438129811604\\
0.1	0.000444647005858\\
0.0562341325190349	0.000416182945261\\
0.0316227766016838	0.000453651897777\\
0.0177827941003892	0.00043585162413\\
0.01	0.000433229262037\\
0.00562341325190349	0.000439455710277\\
0.00316227766016838	0.000430339713735\\
0.00177827941003892	0.000434429144861\\
0.001	0.000428659060842\\
0.000562341325190349	0.0004209690999\\
0.000316227766016838	0.000414857911165\\
0.000177827941003892	0.000412456176531\\
0.0001	0.000421787506363\\
};
\addlegendentry{MWPM~\cite{fowler2013minimum}}

\addplot [color=mycolor1, line width=1.0pt, mark size=2.8pt, mark=asterisk, mark options={solid, mycolor1}]
  table[row sep=crcr]{%
1	1\\
0.562341325190349	1\\
0.316227766016838	0.994038748137\\
0.177827941003892	0.806517311609\\
0.1	0.386398763524\\
0.0562341325190349	0.139984148999\\
0.0316227766016838	0.04196730295\\
0.0177827941003892	0.0125704246683\\
0.01	0.00409125914517\\
0.00562341325190349	0.0011625772502\\
0.00316227766016838	0.000338167194736\\
0.00177827941003892	0.000130655879579\\
0.001	4.51416930031e-05\\
0.000562341325190349	9.50049858805e-06\\
0.000316227766016838	4.11785244277e-06\\
0.000177827941003892	1.56021321499e-06\\
};
\addlegendentry{original BP~\cite{ching2021log}}

\addplot [color=blue, line width=1.0pt, mark size=2.8pt, mark=+, mark options={solid, blue}]
  table[row sep=crcr]{%
1	0.000445737032804\\
0.562341325190349	0.000471504618388\\
0.316227766016838	0.000474667021085\\
0.177827941003892	0.000510294303398\\
0.1	0.000513835007579\\
0.0562341325190349	0.00044406535235\\
0.0316227766016838	0.000443224629652\\
0.0177827941003892	0.000439512988765\\
0.01	0.000446760361083\\
0.00562341325190349	0.000425471564317\\
0.00316227766016838	0.000450036837693\\
0.00177827941003892	0.000429748632862\\
0.001	0.000442769829143\\
0.000562341325190349	0.000437922358942\\
0.000316227766016838	0.000441306266549\\
0.000177827941003892	0.000448256803276\\
0.0001	0.000436537352772\\
};
\addlegendentry{original BP-OSD~\cite{panteleev2021degenerate}}

\addplot [color=green, line width=1.0pt, mark size=1.8pt, mark=triangle, mark options={solid, rotate=180, green}]
  table[row sep=crcr]{%
1	0.989514348786\\
0.562341325190349	0.648280802292\\
0.316227766016838	0.625523012552\\
0.177827941003892	0.383904295813\\
0.1	0.14297188755\\
0.0562341325190349	0.0479910714286\\
0.0316227766016838	0.0163646990064\\
0.0177827941003892	0.00490454593397\\
0.01	0.00186740495745\\
0.00562341325190349	0.000549007211958\\
0.00316227766016838	0.000154360292358\\
0.00177827941003892	5.55112312282e-05\\
0.001	2.4977926061e-05\\
0.000562341325190349	5.48518489955e-06\\
0.000316227766016838	3.74301352489e-06\\
0.000177827941003892	1.33148026786e-06\\
};
\addlegendentry{enhanced BP}

\addplot [color=red, line width=1.0pt, mark size=2.8pt, mark=o, mark options={solid, red}]
  table[row sep=crcr]{%
1	0.000444736982549\\
0.562341325190349	0.000438185212125\\
0.316227766016838	0.000452112268519\\
0.177827941003892	0.000443994529987\\
0.1	0.000456077460196\\
0.0562341325190349	0.00045483696369\\
0.0316227766016838	0.000440227379579\\
0.0177827941003892	0.000412816100645\\
0.01	0.000209516480284\\
0.00562341325190349	2.90712970653e-05\\
0.00316227766016838	1.57129063456e-05\\
0.00177827941003892	6.37703451315e-06\\
0.001	2.94183465683e-06\\
0.000562341325190349	1.44199739249e-06\\
};
\addlegendentry{extended BP-OSD}

\end{axis}
\end{tikzpicture}%

%% file: Figures/plot-EqNoise.tex
%
%
\definecolor{mycolor1}{rgb}{1.00000,0.00000,1.00000}%
\begin{tikzpicture}[%
thick,scale=1.00, every node/.style={scale=1.00}
]

\begin{axis}[%
width=9.066666699999999cm,
height=6.6115258cm,
at={(0cm,0cm)},
scale only axis,
xmode=log,
xmin=0.0001,
xmax=1,
xminorticks=true,
xlabel style={font=\color{white!15!black}},
xlabel={Physical error rate},
ymode=log,
ymin=1e-06,
ymax=1,
yminorticks=true,
ylabel style={font=\color{white!15!black}},
ylabel={Logical error rate},
axis background/.style={fill=white},
xmajorgrids,
xminorgrids,
ymajorgrids,
yminorgrids,
legend style={at={(0.03,0.97)}, anchor=north west, legend cell align=left, align=left, draw=white!15!black}
]
\addplot [color=black, line width=1.0pt, mark size=2.8pt, mark=x, mark options={solid, black}]
  table[row sep=crcr]{%
1	0.748044882693\\
0.562341325190349	0.745762711864\\
0.316227766016838	0.65127294257\\
0.177827941003892	0.343267280387\\
0.1	0.101373145332\\
0.0562341325190349	0.0227830535505\\
0.0316227766016838	0.004910232901\\
0.0177827941003892	0.00129354912675\\
0.01	0.000603601622929\\
0.00562341325190349	0.000444402598625\\
0.00316227766016838	0.000453428066073\\
0.00177827941003892	0.000425455845549\\
0.001	0.000415869302083\\
0.000562341325190349	0.000416483937787\\
0.000316227766016838	0.000432208473522\\
0.000177827941003892	0.000447990075297\\
0.0001	0.000416280739731\\
};
\addlegendentry{MWPM~\cite{fowler2013minimum}}

\addplot [color=mycolor1, line width=1.0pt, mark size=2.8pt, mark=asterisk, mark options={solid, mycolor1}]
  table[row sep=crcr]{%
1	1\\
0.562341325190349	1\\
0.316227766016838	0.996538081108\\
0.177827941003892	0.811111111111\\
0.1	0.388595564942\\
0.0562341325190349	0.137507673419\\
0.0316227766016838	0.0424571703983\\
0.0177827941003892	0.0122368529114\\
0.01	0.0039304378376\\
0.00562341325190349	0.00113045444269\\
0.00316227766016838	0.000338685142525\\
0.00177827941003892	0.000130275167208\\
0.001	4.77380788924e-05\\
0.000562341325190349	9.91716710721e-06\\
0.000316227766016838	4.08690041748e-06\\
0.000177827941003892	1.54543735738e-06\\
};
\addlegendentry{original BP~\cite{ching2021log}}

\addplot [color=blue, line width=1.0pt, mark size=2.8pt, mark=+, mark options={solid, blue}]
  table[row sep=crcr]{%
1	0.748113755078\\
0.562341325190349	0.757575757576\\
0.316227766016838	0.628036086051\\
0.177827941003892	0.294983642312\\
0.1	0.0708379272326\\
0.0562341325190349	0.0105175754221\\
0.0316227766016838	0.00211675921097\\
0.0177827941003892	0.000775409189891\\
0.01	0.000483363576574\\
0.00562341325190349	0.0004531606836\\
0.00316227766016838	0.00044444638103\\
0.00177827941003892	0.00044759513528\\
0.001	0.000462328350461\\
0.000562341325190349	0.000439675688758\\
0.000316227766016838	0.0004496075048\\
0.000177827941003892	0.000445315637338\\
0.0001	0.000434800691144\\
};
\addlegendentry{original BP-OSD~\cite{panteleev2021degenerate}}

\addplot [color=green, line width=1.0pt, mark size=1.8pt, mark=triangle, mark options={solid, rotate=180, green}]
  table[row sep=crcr]{%
1	0.995233050847\\
0.562341325190349	0.912075029308\\
0.316227766016838	0.841690544413\\
0.177827941003892	0.486683848797\\
0.1	0.168985059248\\
0.0562341325190349	0.0545980087785\\
0.0316227766016838	0.0172221533887\\
0.0177827941003892	0.00471296064176\\
0.01	0.0018419542833\\
0.00562341325190349	0.000536888537644\\
0.00316227766016838	0.000159104156813\\
0.00177827941003892	5.52751875211e-05\\
0.001	2.59400041093e-05\\
0.000562341325190349	5.51844341046e-06\\
0.000316227766016838	3.75875922456e-06\\
0.000177827941003892	1.2565704656e-06\\
};
\addlegendentry{enhanced BP}

\addplot [color=red, line width=1.0pt, mark size=2.8pt, mark=o, mark options={solid, red}]
  table[row sep=crcr]{%
1	0.770850202429\\
0.562341325190349	0.759700476515\\
0.316227766016838	0.629662522202\\
0.177827941003892	0.276315789474\\
0.1	0.0567149521237\\
0.0562341325190349	0.0096975178374\\
0.0316227766016838	0.00205761316872\\
0.0177827941003892	0.0007362330927\\
0.01	0.000256635604074\\
0.00562341325190349	3.61632703627e-05\\
0.00316227766016838	1.72384101428e-05\\
0.00177827941003892	6.25361067847e-06\\
0.001	2.94772378229e-06\\
0.000562341325190349	1.44330878655e-06\\
};
\addlegendentry{extended BP-OSD}

\end{axis}
\end{tikzpicture}%

%% file: Figures/plot-EqNoise-ms.tex
%
%
\definecolor{mycolor1}{rgb}{1.00000,0.00000,1.00000}%
\begin{tikzpicture}[%
thick,scale=1.00, every node/.style={scale=1.00}
]

\begin{axis}[%
width=9.066666699999999cm,
height=6.6115258cm,
at={(0cm,0cm)},
scale only axis,
xmode=log,
xmin=0.0001,
xmax=1,
xminorticks=true,
xlabel style={font=\color{white!15!black}},
xlabel={Physical error rate},
ymode=log,
ymin=0.001,
ymax=1,
yminorticks=true,
ylabel style={font=\color{white!15!black}},
ylabel={Average decoding time~(ms)},
axis background/.style={fill=white},
xmajorgrids,
xminorgrids,
ymajorgrids,
yminorgrids,
legend style={at={(0.03,0.97)}, anchor=north west, legend cell align=left, align=left, draw=white!15!black}
]
\addplot [color=black, line width=1.0pt, mark size=2.8pt, mark=x, mark options={solid, black}]
  table[row sep=crcr]{%
1	0.02697992\\
0.562341325190349	0.02956152\\
0.316227766016838	0.02828016\\
0.177827941003892	0.01847392\\
0.1	0.00823056\\
0.0562341325190349	0.00466108\\
0.0316227766016838	0.002631424\\
0.0177827941003892	0.001910824\\
0.01	0.001745264\\
0.00562341325190349	0.00165312\\
0.00316227766016838	0.001604368\\
0.00177827941003892	0.001563208\\
0.001	0.00144528\\
0.000562341325190349	0.00147804\\
0.000316227766016838	0.001562424\\
0.000177827941003892	0.001530504\\
0.0001	0.001546696\\
};
\addlegendentry{MWPM~\cite{fowler2013minimum}}

\addplot [color=mycolor1, line width=1.0pt, mark size=2.8pt, mark=asterisk, mark options={solid, mycolor1}]
  table[row sep=crcr]{%
1	0.2188888\\
0.562341325190349	0.2148176\\
0.316227766016838	0.2161288\\
0.177827941003892	0.1782672\\
0.1	0.0917096\\
0.0562341325190349	0.03962776\\
0.0316227766016838	0.01890088\\
0.0177827941003892	0.01195656\\
0.01	0.00979776\\
0.00562341325190349	0.0089316\\
0.00316227766016838	0.00853048\\
0.00177827941003892	0.00835744\\
0.001	0.00799404\\
0.000562341325190349	0.0080056\\
0.000316227766016838	0.00796824\\
0.000177827941003892	0.00802888\\
};
\addlegendentry{original BP~\cite{ching2021log}}

\addplot [color=blue, line width=1.0pt, mark size=2.8pt, mark=+, mark options={solid, blue}]
  table[row sep=crcr]{%
1	0.2727768\\
0.562341325190349	0.3017544\\
0.316227766016838	0.3055904\\
0.177827941003892	0.2429008\\
0.1	0.1243032\\
0.0562341325190349	0.05194744\\
0.0316227766016838	0.02198696\\
0.0177827941003892	0.01265272\\
0.01	0.00968128\\
0.00562341325190349	0.00866776\\
0.00316227766016838	0.00846304\\
0.00177827941003892	0.00816136\\
0.001	0.007910008\\
0.000562341325190349	0.007769016\\
0.000316227766016838	0.007856064\\
0.000177827941003892	0.007726872\\
0.0001	0.007316408\\
};
\addlegendentry{original BP-OSD~\cite{panteleev2021degenerate}}

\addplot [color=green, line width=1.0pt, mark size=1.8pt, mark=triangle, mark options={solid, rotate=180, green}]
  table[row sep=crcr]{%
1	0.290744\\
0.562341325190349	0.3087352\\
0.316227766016838	0.305076\\
0.177827941003892	0.232016\\
0.1	0.1131504\\
0.0562341325190349	0.04708\\
0.0316227766016838	0.02246048\\
0.0177827941003892	0.0135676\\
0.01	0.00977744\\
0.00562341325190349	0.00866856\\
0.00316227766016838	0.007841096\\
0.00177827941003892	0.007740976\\
0.001	0.007316144\\
0.000562341325190349	0.007148792\\
0.000316227766016838	0.007104584\\
0.000177827941003892	0.0071\\
};
\addlegendentry{enhanced BP}

\addplot [color=red, line width=1.0pt, mark size=2.8pt, mark=o, mark options={solid, red}]
  table[row sep=crcr]{%
1	0.4011752\\
0.562341325190349	0.3854176\\
0.316227766016838	0.3829368\\
0.177827941003892	0.2913472\\
0.1	0.1332704\\
0.0562341325190349	0.0517664\\
0.0316227766016838	0.02138256\\
0.0177827941003892	0.01245576\\
0.01	0.0094716\\
0.00562341325190349	0.008416\\
0.00316227766016838	0.007946368\\
0.00177827941003892	0.007635312\\
0.001	0.007286264\\
0.000562341325190349	0.007204384\\
};
\addlegendentry{extended BP-OSD}

\end{axis}
\end{tikzpicture}%

%% file: PRA_2024.bbl
\begin{thebibliography}{22}%
\makeatletter
\providecommand \@ifxundefined [1]{%
 \@ifx{#1\undefined}
}%
\providecommand \@ifnum [1]{%
 \ifnum #1\expandafter \@firstoftwo
 \else \expandafter \@secondoftwo
 \fi
}%
\providecommand \@ifx [1]{%
 \ifx #1\expandafter \@firstoftwo
 \else \expandafter \@secondoftwo
 \fi
}%
\providecommand \natexlab [1]{#1}%
\providecommand \enquote  [1]{``#1''}%
\providecommand \bibnamefont  [1]{#1}%
\providecommand \bibfnamefont [1]{#1}%
\providecommand \citenamefont [1]{#1}%
\providecommand \href@noop [0]{\@secondoftwo}%
\providecommand \href [0]{\begingroup \@sanitize@url \@href}%
\providecommand \@href[1]{\@@startlink{#1}\@@href}%
\providecommand \@@href[1]{\endgroup#1\@@endlink}%
\providecommand \@sanitize@url [0]{\catcode `\\12\catcode `\$12\catcode `\&12\catcode `\#12\catcode `\^12\catcode `\_12\catcode `\%12\relax}%
\providecommand \@@startlink[1]{}%
\providecommand \@@endlink[0]{}%
\providecommand \url  [0]{\begingroup\@sanitize@url \@url }%
\providecommand \@url [1]{\endgroup\@href {#1}{\urlprefix }}%
\providecommand \urlprefix  [0]{URL }%
\providecommand \Eprint [0]{\href }%
\providecommand \doibase [0]{https://doi.org/}%
\providecommand \selectlanguage [0]{\@gobble}%
\providecommand \bibinfo  [0]{\@secondoftwo}%
\providecommand \bibfield  [0]{\@secondoftwo}%
\providecommand \translation [1]{[#1]}%
\providecommand \BibitemOpen [0]{}%
\providecommand \bibitemStop [0]{}%
\providecommand \bibitemNoStop [0]{.\EOS\space}%
\providecommand \EOS [0]{\spacefactor3000\relax}%
\providecommand \BibitemShut  [1]{\csname bibitem#1\endcsname}%
\let\auto@bib@innerbib\@empty
\bibitem [{\citenamefont {Gottesman}(1997)}]{gottesman1997stabilizer}%
  \BibitemOpen
  \bibfield  {author} {\bibinfo {author} {\bibfnamefont {D.}~\bibnamefont {Gottesman}},\ }\href@noop {} {\emph {\bibinfo {title} {Stabilizer codes and quantum error correction}}}\ (\bibinfo  {publisher} {California Institute of Technology},\ \bibinfo {address} {Pasadena, CA},\ \bibinfo {year} {1997})\BibitemShut {NoStop}%
\bibitem [{\citenamefont {Nielsen}\ and\ \citenamefont {Chuang}(2010)}]{nielsen2010quantum}%
  \BibitemOpen
  \bibfield  {author} {\bibinfo {author} {\bibfnamefont {M.~A.}\ \bibnamefont {Nielsen}}\ and\ \bibinfo {author} {\bibfnamefont {I.~L.}\ \bibnamefont {Chuang}},\ }\href@noop {} {\emph {\bibinfo {title} {Quantum computation and quantum information}}}\ (\bibinfo  {publisher} {Cambridge University Press},\ \bibinfo {address} {Cambridge},\ \bibinfo {year} {2010})\BibitemShut {NoStop}%
\bibitem [{\citenamefont {Dennis}\ \emph {et~al.}(2002)\citenamefont {Dennis}, \citenamefont {Kitaev}, \citenamefont {Landahl},\ and\ \citenamefont {Preskill}}]{dennis2002topological}%
  \BibitemOpen
  \bibfield  {author} {\bibinfo {author} {\bibfnamefont {E.}~\bibnamefont {Dennis}}, \bibinfo {author} {\bibfnamefont {A.}~\bibnamefont {Kitaev}}, \bibinfo {author} {\bibfnamefont {A.}~\bibnamefont {Landahl}},\ and\ \bibinfo {author} {\bibfnamefont {J.}~\bibnamefont {Preskill}},\ }\bibfield  {title} {\bibinfo {title} {Topological quantum memory},\ }\href {https://doi.org/10.1063/1.1499754} {\bibfield  {journal} {\bibinfo  {journal} {Journal of Mathematical Physics}\ }\textbf {\bibinfo {volume} {43}},\ \bibinfo {pages} {4452} (\bibinfo {year} {2002})}\BibitemShut {NoStop}%
\bibitem [{\citenamefont {Kitaev}(2003)}]{kitaev2003fault}%
  \BibitemOpen
  \bibfield  {author} {\bibinfo {author} {\bibfnamefont {A.~Y.}\ \bibnamefont {Kitaev}},\ }\bibfield  {title} {\bibinfo {title} {Fault-tolerant quantum computation by anyons},\ }\href {https://doi.org/10.1016/S0003-4916(02)00018-0} {\bibfield  {journal} {\bibinfo  {journal} {Annals of physics}\ }\textbf {\bibinfo {volume} {303}},\ \bibinfo {pages} {2} (\bibinfo {year} {2003})}\BibitemShut {NoStop}%
\bibitem [{\citenamefont {Kolmogorov}(2009)}]{kolmogorov2009blossom}%
  \BibitemOpen
  \bibfield  {author} {\bibinfo {author} {\bibfnamefont {V.}~\bibnamefont {Kolmogorov}},\ }\bibfield  {title} {\bibinfo {title} {Blossom {V}: a new implementation of a minimum cost perfect matching algorithm},\ }\href {https://doi.org/10.1007/s12532-009-0002-8} {\bibfield  {journal} {\bibinfo  {journal} {Mathematical Programming Computation}\ }\textbf {\bibinfo {volume} {1}},\ \bibinfo {pages} {43} (\bibinfo {year} {2009})}\BibitemShut {NoStop}%
\bibitem [{\citenamefont {Fowler}(2014)}]{fowler2013minimum}%
  \BibitemOpen
  \bibfield  {author} {\bibinfo {author} {\bibfnamefont {A.~G.}\ \bibnamefont {Fowler}},\ }\href {https://arxiv.org/abs/1307.1740} {\bibinfo {title} {Minimum weight perfect matching of fault-tolerant topological quantum error correction in average {$O(1)$} parallel time}} (\bibinfo {year} {2014}),\ \Eprint {https://arxiv.org/abs/1307.1740} {arXiv:1307.1740 [quant-ph]} \BibitemShut {NoStop}%
\bibitem [{\citenamefont {Old}\ and\ \citenamefont {Rispler}(2023)}]{old2023generalized}%
  \BibitemOpen
  \bibfield  {author} {\bibinfo {author} {\bibfnamefont {J.}~\bibnamefont {Old}}\ and\ \bibinfo {author} {\bibfnamefont {M.}~\bibnamefont {Rispler}},\ }\bibfield  {title} {\bibinfo {title} {Generalized belief propagation algorithms for decoding of surface codes},\ }\href {https://doi.org/10.22331/q-2023-06-07-1037} {\bibfield  {journal} {\bibinfo  {journal} {{Quantum}}\ }\textbf {\bibinfo {volume} {7}},\ \bibinfo {pages} {1037} (\bibinfo {year} {2023})}\BibitemShut {NoStop}%
\bibitem [{\citenamefont {Chytas}\ \emph {et~al.}(2024)\citenamefont {Chytas}, \citenamefont {Pacenti}, \citenamefont {Raveendran}, \citenamefont {Flanagan},\ and\ \citenamefont {Vasić}}]{chytas2024enhanced}%
  \BibitemOpen
  \bibfield  {author} {\bibinfo {author} {\bibfnamefont {D.}~\bibnamefont {Chytas}}, \bibinfo {author} {\bibfnamefont {M.}~\bibnamefont {Pacenti}}, \bibinfo {author} {\bibfnamefont {N.}~\bibnamefont {Raveendran}}, \bibinfo {author} {\bibfnamefont {M.~F.}\ \bibnamefont {Flanagan}},\ and\ \bibinfo {author} {\bibfnamefont {B.}~\bibnamefont {Vasić}},\ }\bibfield  {title} {\bibinfo {title} {Enhanced message-passing decoding of degenerate quantum codes utilizing trapping set dynamics},\ }\href {https://doi.org/10.1109/LCOMM.2024.3356312} {\bibfield  {journal} {\bibinfo  {journal} {IEEE Communications Letters}\ }\textbf {\bibinfo {volume} {28}},\ \bibinfo {pages} {444} (\bibinfo {year} {2024})}\BibitemShut {NoStop}%
\bibitem [{\citenamefont {Huang}\ \emph {et~al.}(2023)\citenamefont {Huang}, \citenamefont {Hu},\ and\ \citenamefont {Ueng}}]{huang2023branch}%
  \BibitemOpen
  \bibfield  {author} {\bibinfo {author} {\bibfnamefont {T.-H.}\ \bibnamefont {Huang}}, \bibinfo {author} {\bibfnamefont {T.-A.}\ \bibnamefont {Hu}},\ and\ \bibinfo {author} {\bibfnamefont {Y.-L.}\ \bibnamefont {Ueng}},\ }\bibfield  {title} {\bibinfo {title} {Branch-assisted sign-flipping belief propagation decoding for topological quantum codes based on hypergraph product structure},\ }\href {https://doi.org/10.1109/TQE.2023.3279379} {\bibfield  {journal} {\bibinfo  {journal} {IEEE Transactions on Quantum Engineering}\ }\textbf {\bibinfo {volume} {4}},\ \bibinfo {pages} {1} (\bibinfo {year} {2023})}\BibitemShut {NoStop}%
\bibitem [{\citenamefont {Panteleev}\ and\ \citenamefont {Kalachev}(2021)}]{panteleev2021degenerate}%
  \BibitemOpen
  \bibfield  {author} {\bibinfo {author} {\bibfnamefont {P.}~\bibnamefont {Panteleev}}\ and\ \bibinfo {author} {\bibfnamefont {G.}~\bibnamefont {Kalachev}},\ }\bibfield  {title} {\bibinfo {title} {Degenerate quantum {LDPC} codes with good finite length performance},\ }\href {https://doi.org/10.22331/q-2021-11-22-585} {\bibfield  {journal} {\bibinfo  {journal} {{Quantum}}\ }\textbf {\bibinfo {volume} {5}},\ \bibinfo {pages} {585} (\bibinfo {year} {2021})}\BibitemShut {NoStop}%
\bibitem [{\citenamefont {Liang}\ \emph {et~al.}(2024)\citenamefont {Liang}, \citenamefont {Wang}, \citenamefont {Li},\ and\ \citenamefont {Ma}}]{liang2024BPLCOSD}%
  \BibitemOpen
  \bibfield  {author} {\bibinfo {author} {\bibfnamefont {J.}~\bibnamefont {Liang}}, \bibinfo {author} {\bibfnamefont {Q.}~\bibnamefont {Wang}}, \bibinfo {author} {\bibfnamefont {L.}~\bibnamefont {Li}},\ and\ \bibinfo {author} {\bibfnamefont {X.}~\bibnamefont {Ma}},\ }\bibfield  {title} {\bibinfo {title} {The {BP-LCOSD} algorithm for toric codes},\ }in\ \href {https://doi.org/10.1109/ISIT-W61686.2024.10591758} {\emph {\bibinfo {booktitle} {2024 IEEE International Symposium on Information Theory Workshops (ISIT-W)}}}\ (\bibinfo {year} {2024})\ pp.\ \bibinfo {pages} {1--6}\BibitemShut {NoStop}%
\bibitem [{\citenamefont {Meinerz}\ \emph {et~al.}(2022)\citenamefont {Meinerz}, \citenamefont {Park},\ and\ \citenamefont {Trebst}}]{meinerz2022scalable}%
  \BibitemOpen
  \bibfield  {author} {\bibinfo {author} {\bibfnamefont {K.}~\bibnamefont {Meinerz}}, \bibinfo {author} {\bibfnamefont {C.-Y.}\ \bibnamefont {Park}},\ and\ \bibinfo {author} {\bibfnamefont {S.}~\bibnamefont {Trebst}},\ }\bibfield  {title} {\bibinfo {title} {Scalable neural decoder for topological surface codes},\ }\href@noop {} {\bibfield  {journal} {\bibinfo  {journal} {Physical Review Letters}\ }\textbf {\bibinfo {volume} {128}},\ \bibinfo {pages} {080505} (\bibinfo {year} {2022})}\BibitemShut {NoStop}%
\bibitem [{\citenamefont {Nemec}(2023)}]{nemec2023quantum}%
  \BibitemOpen
  \bibfield  {author} {\bibinfo {author} {\bibfnamefont {A.}~\bibnamefont {Nemec}},\ }\bibfield  {title} {\bibinfo {title} {Quantum data-syndrome codes: Subsystem and impure code constructions},\ }\href {https://doi.org/10.1007/s11128-023-04166-z} {\bibfield  {journal} {\bibinfo  {journal} {Quantum Information Processing}\ }\textbf {\bibinfo {volume} {22}},\ \bibinfo {pages} {408} (\bibinfo {year} {2023})}\BibitemShut {NoStop}%
\bibitem [{\citenamefont {Bonilla~Ataides}\ \emph {et~al.}(2021)\citenamefont {Bonilla~Ataides}, \citenamefont {Tuckett}, \citenamefont {Bartlett}, \citenamefont {Flammia},\ and\ \citenamefont {Brown}}]{bonilla2021xzzx}%
  \BibitemOpen
  \bibfield  {author} {\bibinfo {author} {\bibfnamefont {J.~P.}\ \bibnamefont {Bonilla~Ataides}}, \bibinfo {author} {\bibfnamefont {D.~K.}\ \bibnamefont {Tuckett}}, \bibinfo {author} {\bibfnamefont {S.~D.}\ \bibnamefont {Bartlett}}, \bibinfo {author} {\bibfnamefont {S.~T.}\ \bibnamefont {Flammia}},\ and\ \bibinfo {author} {\bibfnamefont {B.~J.}\ \bibnamefont {Brown}},\ }\bibfield  {title} {\bibinfo {title} {The {XZZX} surface code},\ }\href {https://doi.org/10.1038/s41467-021-22274-1} {\bibfield  {journal} {\bibinfo  {journal} {Nature communications}\ }\textbf {\bibinfo {volume} {12}},\ \bibinfo {pages} {2172} (\bibinfo {year} {2021})}\BibitemShut {NoStop}%
\bibitem [{\citenamefont {Lidar}\ and\ \citenamefont {Brun}(2013)}]{lidar2013quantum}%
  \BibitemOpen
  \bibfield  {author} {\bibinfo {author} {\bibfnamefont {D.~A.}\ \bibnamefont {Lidar}}\ and\ \bibinfo {author} {\bibfnamefont {T.~A.}\ \bibnamefont {Brun}},\ }\href@noop {} {\emph {\bibinfo {title} {Quantum error correction}}}\ (\bibinfo  {publisher} {Cambridge University Press},\ \bibinfo {address} {Cambridge},\ \bibinfo {year} {2013})\BibitemShut {NoStop}%
\bibitem [{\citenamefont {Kschischang}\ \emph {et~al.}(2001)\citenamefont {Kschischang}, \citenamefont {Frey},\ and\ \citenamefont {Loeliger}}]{kschischang2001factor}%
  \BibitemOpen
  \bibfield  {author} {\bibinfo {author} {\bibfnamefont {F.}~\bibnamefont {Kschischang}}, \bibinfo {author} {\bibfnamefont {B.}~\bibnamefont {Frey}},\ and\ \bibinfo {author} {\bibfnamefont {H.-A.}\ \bibnamefont {Loeliger}},\ }\bibfield  {title} {\bibinfo {title} {Factor graphs and the sum-product algorithm},\ }\href {https://doi.org/10.1109/18.910572} {\bibfield  {journal} {\bibinfo  {journal} {IEEE Transactions on Information Theory}\ }\textbf {\bibinfo {volume} {47}},\ \bibinfo {pages} {498} (\bibinfo {year} {2001})}\BibitemShut {NoStop}%
\bibitem [{\citenamefont {Fossorier}\ and\ \citenamefont {Lin}(1995)}]{fossorier1995soft}%
  \BibitemOpen
  \bibfield  {author} {\bibinfo {author} {\bibfnamefont {M.~P.}\ \bibnamefont {Fossorier}}\ and\ \bibinfo {author} {\bibfnamefont {S.}~\bibnamefont {Lin}},\ }\bibfield  {title} {\bibinfo {title} {Soft-decision decoding of linear block codes based on ordered statistics},\ }\href@noop {} {\bibfield  {journal} {\bibinfo  {journal} {IEEE Transactions on Information Theory}\ }\textbf {\bibinfo {volume} {41}},\ \bibinfo {pages} {1379} (\bibinfo {year} {1995})}\BibitemShut {NoStop}%
\bibitem [{\citenamefont {Kuo}\ and\ \citenamefont {Lai}(2022)}]{kuo2022exploiting}%
  \BibitemOpen
  \bibfield  {author} {\bibinfo {author} {\bibfnamefont {K.-Y.}\ \bibnamefont {Kuo}}\ and\ \bibinfo {author} {\bibfnamefont {C.-Y.}\ \bibnamefont {Lai}},\ }\bibfield  {title} {\bibinfo {title} {Exploiting degeneracy in belief propagation decoding of quantum codes},\ }\href {https://doi.org/10.1038/s41534-022-00623-2} {\bibfield  {journal} {\bibinfo  {journal} {npj Quantum Information}\ }\textbf {\bibinfo {volume} {8}},\ \bibinfo {pages} {111} (\bibinfo {year} {2022})}\BibitemShut {NoStop}%
\bibitem [{\citenamefont {Laflamme}\ \emph {et~al.}(1996)\citenamefont {Laflamme}, \citenamefont {Miquel}, \citenamefont {Paz},\ and\ \citenamefont {Zurek}}]{laflamme1996perfect}%
  \BibitemOpen
  \bibfield  {author} {\bibinfo {author} {\bibfnamefont {R.}~\bibnamefont {Laflamme}}, \bibinfo {author} {\bibfnamefont {C.}~\bibnamefont {Miquel}}, \bibinfo {author} {\bibfnamefont {J.~P.}\ \bibnamefont {Paz}},\ and\ \bibinfo {author} {\bibfnamefont {W.~H.}\ \bibnamefont {Zurek}},\ }\bibfield  {title} {\bibinfo {title} {Perfect quantum error correcting code},\ }\href {https://doi.org/10.1103/PhysRevLett.77.198} {\bibfield  {journal} {\bibinfo  {journal} {Phys. Rev. Lett.}\ }\textbf {\bibinfo {volume} {77}},\ \bibinfo {pages} {198} (\bibinfo {year} {1996})}\BibitemShut {NoStop}%
\bibitem [{\citenamefont {Poulin}\ and\ \citenamefont {Chung}(2008)}]{poulin2008iterative}%
  \BibitemOpen
  \bibfield  {author} {\bibinfo {author} {\bibfnamefont {D.}~\bibnamefont {Poulin}}\ and\ \bibinfo {author} {\bibfnamefont {Y.}~\bibnamefont {Chung}},\ }\href {https://arxiv.org/abs/0801.1241} {\bibinfo {title} {On the iterative decoding of sparse quantum codes}} (\bibinfo {year} {2008}),\ \Eprint {https://arxiv.org/abs/0801.1241} {arXiv:0801.1241 [quant-ph]} \BibitemShut {NoStop}%
\bibitem [{\citenamefont {Liang}\ and\ \citenamefont {Ma}(2024)}]{liang2024random}%
  \BibitemOpen
  \bibfield  {author} {\bibinfo {author} {\bibfnamefont {J.}~\bibnamefont {Liang}}\ and\ \bibinfo {author} {\bibfnamefont {X.}~\bibnamefont {Ma}},\ }\href {https://arxiv.org/abs/2401.16709} {\bibinfo {title} {A random coding approach to performance analysis of the ordered statistic decoding with local constraints}} (\bibinfo {year} {2024}),\ \Eprint {https://arxiv.org/abs/2401.16709} {arXiv:2401.16709 [cs.IT]} \BibitemShut {NoStop}%
\bibitem [{\citenamefont {Lai}\ and\ \citenamefont {Kuo}(2021)}]{ching2021log}%
  \BibitemOpen
  \bibfield  {author} {\bibinfo {author} {\bibfnamefont {C.-Y.}\ \bibnamefont {Lai}}\ and\ \bibinfo {author} {\bibfnamefont {K.-Y.}\ \bibnamefont {Kuo}},\ }\bibfield  {title} {\bibinfo {title} {Log-domain decoding of quantum {LDPC} codes over binary finite fields},\ }\href {https://doi.org/10.1109/TQE.2021.3113936} {\bibfield  {journal} {\bibinfo  {journal} {IEEE Transactions on Quantum Engineering}\ }\textbf {\bibinfo {volume} {2}},\ \bibinfo {pages} {1} (\bibinfo {year} {2021})}\BibitemShut {NoStop}%
\end{thebibliography}%
